\documentclass[11pt,a4paper]{article}
\pdfoutput=1
\usepackage{jcappub}
\usepackage{aas_macros_JCAP}

\newcommand{\hMpc}{h^{-1}{\rm Mpc}}

\newcommand{\void}{\mathrm{v}}

\newcommand{\gal}{\mathrm{g}}

\newcommand{\xibar}{\overline{\xi}}
\newcommand{\C}{\mathbf{C}}

\title{Multipole analysis of redshift-space distortions around cosmic voids}

\author[a]{Nico Hamaus,}
\author[b]{Marie-Claude Cousinou,}
\author[b]{Alice Pisani,}
\author[b]{\\Marie Aubert,}
\author[b]{St\'ephanie Escoffier}
\author[a,c,d]{and Jochen Weller}

\affiliation[a]{Universit\"ats-Sternwarte M\"unchen, Fakult\"at f\"ur Physik, Ludwig-Maximilians Universit\"at, Scheinerstr. 1, D-81679 M\"unchen, Germany}
\affiliation[b]{Aix Marseille Univ, CNRS/IN2P3, CPPM, \\163 avenue de Luminy, F-13288, Marseille, France}
\affiliation[c]{Max Planck Institute for Extraterrestrial Physics, \\Giessenbachstr. 1, D-85748 Garching, Germany}
\affiliation[d]{Excellence Cluster Universe, \\Bolzmannstr. 2, D-85748 Garching, Germany}

\emailAdd{hamaus@usm.lmu.de}
\emailAdd{cousinou@cppm.in2p3.fr}
\emailAdd{pisani@cppm.in2p3.fr}
\emailAdd{maubert@cppm.in2p3.fr}
\emailAdd{escoffier@cppm.in2p3.fr}
\emailAdd{jochen.weller@usm.lmu.de}

\abstract{
We perform a comprehensive redshift-space distortion analysis based on cosmic voids in the large-scale distribution of galaxies observed with the Sloan Digital Sky Survey. To this end, we measure multipoles of the void-galaxy cross-correlation function and compare them with standard model predictions in cosmology. Merely considering linear-order theory allows us to accurately describe the data on the entire available range of scales and to probe void-centric distances down to about $2\hMpc$. Common systematics, such as the Fingers-of-God effect, scale-dependent galaxy bias, and nonlinear clustering do not seem to play a significant role in our analysis. We constrain the growth rate of structure via the redshift-space distortion parameter $\beta$ at two median redshifts, $\beta(\bar{z}=0.32)=0.599^{+0.134}_{-0.124}$ and $\beta(\bar{z}=0.54)=0.457^{+0.056}_{-0.054}$, with a precision that is competitive with state-of-the-art galaxy-clustering results. While the high-redshift constraint perfectly agrees with model expectations, we observe a mild $2\sigma$ deviation at $\bar{z}=0.32$, which increases to $3\sigma$ when the data is restricted to the lowest available redshift range of $0.15<z<0.33$. 
}

\date{\today}


\arxivnumber{1705.05328}

\begin{document}
\maketitle

\section{Introduction\label{sec:intro}}
Extensive ongoing and planned galaxy redshift surveys allow us to map out ever larger portions of large-scale structure in the observable Universe~\cite[e.g.,][]{DES,EUCLID,LSST,SDSS,WFIRST}. Although these maps only contain snapshots of structure formation at each observed epoch, redshift-space distortions (RSD) in the distribution of galaxies can reveal dynamical information from the cosmic web, such as the growth rate of structure~\cite{Peacock2001,Guzzo2008}. The latter is tightly linked to the nature of gravity, which constitutes the only relevant interaction beyond Megaparsec (Mpc) scales, and therefore governs the buildup of structures in the Universe. RSDs have been analyzed extensively in the two-point statistics of galaxies, i.e., their correlation function~\cite[e.g.,][]{Satpathy2017,SanchezA2017} and their power spectrum~\cite[e.g.,][]{Beutler2017,Grieb2017}. Although both approaches are equivalent in principle, the treatment of systematics to model the data differs quite a lot. A major challenge is posed by the complexity of nonlinear interactions of galaxies on small (intra-cluster) scales, where perturbative methods fail~\cite[e.g.,][]{Scoccimarro2004,Matsubara2008,Taruya2010,Okumura2015}. Most commonly, effective models are evoked in describing the nonlinear regime~\cite[e.g.,][]{Percival2009,Reid2011,White2015}, but the extractable cosmological information is fundamentally limited through the process of shell crossing and virialization.

One way to circumvent this problem is to focus on regions in the Universe that have hardly undergone any virialization at all: \emph{cosmic voids}, vast near-empty domains of the cosmic web that are dominated by coherent bulk flows~\cite{Shandarin2011,Falck2015,Ramachandra2017}. However, this approach requires large contiguous maps of large-scale structure in order to obtain a sufficient number of voids to create statistically significant sample sizes. Only the current generation of redshift surveys has been able to reach that requirement, which triggered a considerable amount of activity and scientific results in this comparably young field. Cosmic voids have successfully been analyzed for their sensitivity to Alcock-Paczy\'nski (AP) distortions~\cite{Sutter2012b,Sutter2014b,Hamaus2016,Mao2017}, their ability to act as weak gravitational lenses (WL)~\cite{Melchior2014,Clampitt2015,SanchezC2017,Chantavat2017}, to exhibit an integrated Sachs-Wolfe (ISW) signature~\cite{Granett2008,Nadathur2016,Cai2017,Kovacs2017}, to feature baryon acoustic oscillations (BAO)~\cite{Kitaura2016}, and to show a unique RSD pattern~\cite{Paz2013,Hamaus2016,Achitouv2017,Hawken2016}, to highlight only some of the most recent observational discoveries in the cosmological context of voids.

In this paper we explore public data from the Sloan Digital Sky Survey (SDSS) and present the first measurement of the void-galaxy cross-correlation function multipoles in redshift space, enabling us to constrain the growth rate of structure. After a brief recapitulation of theoretical prerequisites in section~\ref{sec:theory}, details about the considered data sets and catalogs are given in section~\ref{sec:data}. Section~\ref{sec:analysis} presents the results of our analysis, followed by a discussion in section~\ref{sec:discussion} and the final conclusions in section~\ref{sec:conclusion}.

\section{Theory \label{sec:theory}}
Let us consider a void center at comoving coordinate $\mathbf{X}$ located on our line of sight, and a galaxy at location $\mathbf{x}$, with separation $\mathbf{r}=\mathbf{x}-\mathbf{X}$ from it. Because we use the redshift $z$ of the galaxy to determine its comoving distance to us, its peculiar velocity $\mathbf{v}$ will have a contribution via the Doppler effect~\cite{Kaiser1987,Hamilton1998}, so the inferred separation between the galaxy and the void center in redshift space is
\begin{equation}
 \mathbf{s} = \mathbf{r} + (1+z)\frac{\hat{\mathbf{X}}\cdot\mathbf{v}}{H(z)}\hat{\mathbf{X}}\;, \label{rsd}
\end{equation}
where $H(z)$ is the Hubble rate, $\hat{\mathbf{X}}=\mathbf{X}/|\mathbf{X}|$, and we assumed $|\mathbf{r}|\ll|\mathbf{X}|$ such that $\mathbf{x}$ and $\mathbf{X}$ are approximately parallel (distant-observer approximation). As long as we only consider relative motions between galaxies and void centers on scales of the void extent, bulk motions between different voids can be neglected~\cite{Sutter2014c,Lambas2016,Chuang2016a}. The galaxy's peculiar velocity is sourced by the underlying mass distribution of the void, which obeys spherical symmetry in the cosmic average. According to the linearized mass-conservation equation~\cite{Peebles1980}, it can be related to the average mass-density contrast $\Delta(r)$ within radius $r=|\mathbf{r}|$ around the void center~\cite{Hamaus2014b},
\begin{equation}
 \mathbf{v}(\mathbf{r}) = -\frac{1}{3}\frac{f(z)H(z)}{1+z}\mathbf{r}\Delta(r)\;, \label{vlin}
\end{equation}
where $f(z)$ is the logarithmic growth rate for linear density perturbations. Assuming General Relativity (GR) and the standard $\Lambda$CDM-model for cosmology, it can be expressed as a power of the matter-density parameter, $f(z)=\Omega_\mathrm{m}^{\;\gamma}(z)$, with a growth index of $\gamma\simeq0.55$~\cite{Linder2005}. Unfortunately, the total mass-density contrast $\Delta(r)$ around voids is not directly observable~\cite{Leclercq2015}, but with the help of simulations it has been demonstrated that its relation to the corresponding average galaxy-density contrast $\xibar(r)$ is remarkably linear~\cite{Pollina2017},
\begin{equation}
 \xibar(r) = b\Delta(r)\;,
\end{equation}
with a single bias parameter $b$. Therefore, in exchanging $\Delta(r)$ with the observable $\xibar(r)$ in equation~(\ref{vlin}), we can absorb the bias parameter into the definition of the growth rate by defining the relative growth rate $\beta\equiv f/b$. Then, plugging this into equation~(\ref{rsd}), we have
\begin{equation}
 \mathbf{s} = \mathbf{r} - \frac{\beta}{3}\hat{\mathbf{X}}\cdot\mathbf{r}\,\xibar(r)\hat{\mathbf{X}}\;. \label{rsd2}
\end{equation}
The total number of galaxies cannot be altered by RSDs, therefore the void-galaxy cross-correlation functions in real and redshift space, $\xi$ and $\xi^s$, must satisfy
\begin{equation}
 \int[1+\xi(r)]\mathrm{d}^3r = \int[1+\xi^s(\mathbf{s})]\mathrm{d}^3s = \int[1+\xi^s(\mathbf{r})]\det\!\left(\frac{\partial\mathbf{s}}{\partial\mathbf{r}}\right)\mathrm{d}^3r \;. \label{number}
\end{equation}
In the last step we introduce the determinant of the Jacobian $\partial\mathbf{s}/\partial\mathbf{{r}}$ to perform a coordinate transformation between $\mathbf{s}$ and $\mathbf{r}$. Using equation~(\ref{rsd2}) it equates to
\begin{equation}
 \det\!\left(\frac{\partial\mathbf{s}}{\partial\mathbf{r}}\right) = 1-\frac{\beta}{3}\xibar(r)-\frac{\beta}{3}\hat{\mathbf{X}}\cdot\mathbf{r}\,\frac{\partial\xibar(r)}{\partial r}\frac{\mathbf{r}}{r}\cdot\hat{\mathbf{X}}\;. \label{jac}
\end{equation}
The average galaxy-density contrast is an integral over the void-galaxy cross-correlation function,
\begin{equation}
 \xibar(r) = \frac{3}{r^3}\int_0^r\xi(r')r'^2\mathrm{d}r'\;,
\end{equation}
with $\partial\xibar(r)/\partial r = 3/r\left[\xi(r)-\xibar(r)\right]$. Defining the angle $\vartheta$ between the line-of-sight direction $\mathbf{X}$ and the separation vector $\mathbf{r}$ via
\begin{equation}
 \cos\vartheta = \frac{\mathbf{X}\cdot\mathbf{r}}{|\mathbf{X}||\mathbf{r}|} \equiv \mu \;,
\end{equation}
the determinant of the Jacobian can be written as
\begin{equation}
 \det\!\left(\frac{\partial\mathbf{s}}{\partial\mathbf{r}}\right) = 1-\frac{\beta}{3}\xibar(r)-\beta\mu^2\left[\xi(r)-\xibar(r)\right]\;.
\end{equation}
Using this in equation~(\ref{number}) and solving for $\xi^s$ to linear order in $\xi$ and $\xibar$ finally yields a relation between the real-space and redshift-space void-galaxy cross-correlation functions~\cite{Cai2016},
\begin{equation}
 \xi^s(r,\mu) = \xi(r) + \frac{\beta}{3}\xibar(r) + \beta\mu^2\left[\xi(r)-\xibar(r)\right]\;. \label{xi_s}
\end{equation}
As $\xi^s$ is no longer isotropic, one can decompose the redshift-space correlation function into multipoles using the Legendre polynomials $P_\ell(\mu)$ via
\begin{equation}
 \xi_\ell(r) = \int_0^1\xi^s(r,\mu)(1+2\ell)P_\ell(\mu)\mathrm{d}\mu\;. \label{xi_l}
\end{equation}
The only non-vanishing multipoles of equation~(\ref{xi_s}) are the monopole with $P_0=1$ and the quadrupole with $P_2=(3\mu^2-1)/2$,
\begin{eqnarray}
 \xi_0(r) &=& \left(1+\frac{\beta}{3}\right)\xi(r)\;, \\
 \xi_2(r) &=& \frac{2\beta}{3}\left[\xi(r)-\xibar(r)\right]\;.
\end{eqnarray}
Hence, these two functions fully determine the void-galaxy cross-correlation function in redshift space, which can then be expressed as
\begin{equation}
 \xi^s(r,\mu) = \xi_0(r) + \frac{3\mu^2-1}{2}\xi_2(r)\;.
\end{equation}
The monopole and quadrupole are related via a simple linear equation,
\begin{equation}
 \xi_0(r) - \xibar_0(r) = \xi_2(r)\frac{3+\beta}{2\beta}\;, \label{xi_0_2}
\end{equation}
which, given the multipole measurements, solely depends on the relative growth rate $\beta=f/b$.

\section{Data \label{sec:data}}

\subsection{Galaxy catalogs}
For the analysis in this paper we make use of public data from the SDSS-III~\cite{Eisenstein2011}, namely Data Release 12 (DR12) of the large-scale structure catalogs from the Baryon Oscillation Spectroscopic Survey (BOSS)~\cite{Dawson2013}. The survey targeted two distinct galaxy samples, denoted as LOWZ and CMASS, in both the northern and southern galactic hemispheres over an area of nearly $10\,000$ square degrees with comoving number density of a few times $10^{-4}\,h^{-3}\mathrm{Mpc}^3$~\cite{Reid2016}. LOWZ contains $361\,762$ galaxies in the redshift range $0.15<z<0.43$ with median $\bar{z}=0.32$, and CMASS features $777\,202$ galaxies in the range $0.43<z<0.70$ with $\bar{z}=0.54$. The linear galaxy bias of both samples is estimated as $b\simeq1.85$~\cite{Alam2016}. We also consider publicly available mock galaxy catalogs produced with the \emph{Quick Particle Mesh} (QPM) method~\cite{White2014}. These mocks were set up to specifically mimic the LOWZ and CMASS samples in BOSS, including various observational effects, adopting the following parameter values of a flat $\Lambda$CDM model: $(\Omega_\mathrm{m},\Omega_\Lambda,\Omega_\mathrm{b},\sigma_8,h,n_s) = (0.29, 0.71, 0.0458, 0.80, 0.7, 0.97)$~\cite{GilMarin2017}. The linear bias of the mock galaxies amounts to $b=2.2$~\cite{Alam2016}. Furthermore, we make use of random catalogs provided by the BOSS collaboration, featuring the same angular and redshift distributions as the observed galaxies, but about $50$ times as many objects~\cite{Reid2016}.

\subsection{Void catalogs}
We make use of the Void IDentification and Examination toolkit \textsc{vide}~\cite{Sutter2015}, an open-source software repository\footnote{\url{https://bitbucket.org/cosmicvoids/vide_public/}} to find voids in simulations and observational data sets. The framework is based on \textsc{zobov}~\cite{Neyrinck2008}, an algorithm that implements the watershed transform~\cite{Platen2007} to identify nested basins around local minima in a three-dimensional density field. The density field is estimated via a Voronoi tessellation of the tracer particles, such that each particle gets assigned to a Voronoi cell whose inverse volume provides a density estimate within the cell. Neighboring basins are merged to form larger voids, if the minimum density on the separating ridge between them falls below $20\%$ of the mean tracer density. This mitigates the probability of identifying spurious voids that can arise from Poisson fluctuations in places where tracer particles are sparse, and prevents highly over-dense structures to be included inside voids. When running \textsc{vide} on the BOSS galaxy and mock samples, their sky positions and redshifts are transformed to comoving coordinates assuming a flat $\Lambda$CDM cosmology with $\Omega_\mathrm{m}=0.31$. At the survey edges, boundary particles are injected to prevent voids from extending outside the surveyed region, and any void that intersects with the survey mask is discarded in the final catalog. Further details on the catalog creation are described in reference~\cite{Sutter2012a}.

\begin{figure*}[!t]
\centering
\resizebox{\hsize}{!}{
\includegraphics{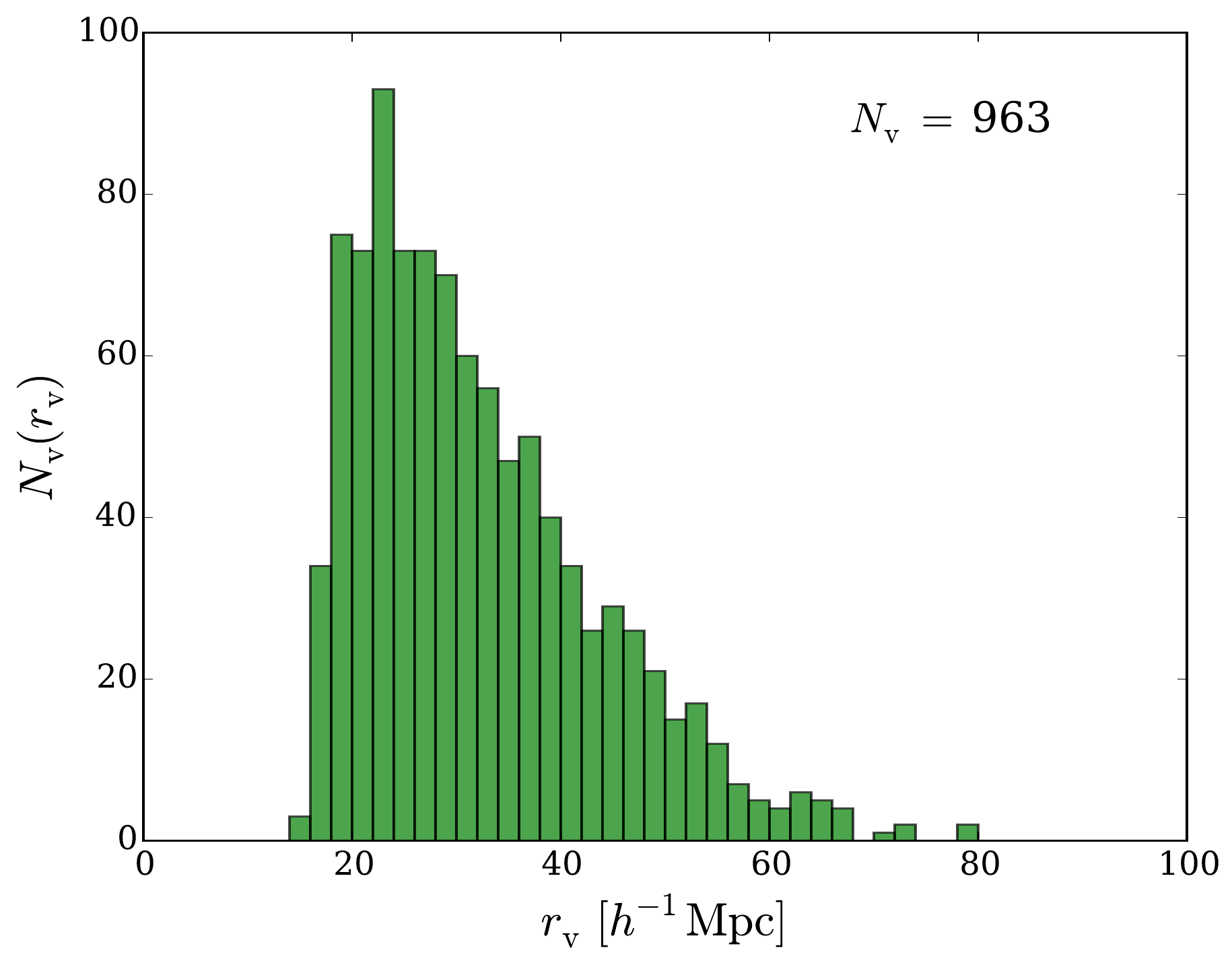}
\includegraphics{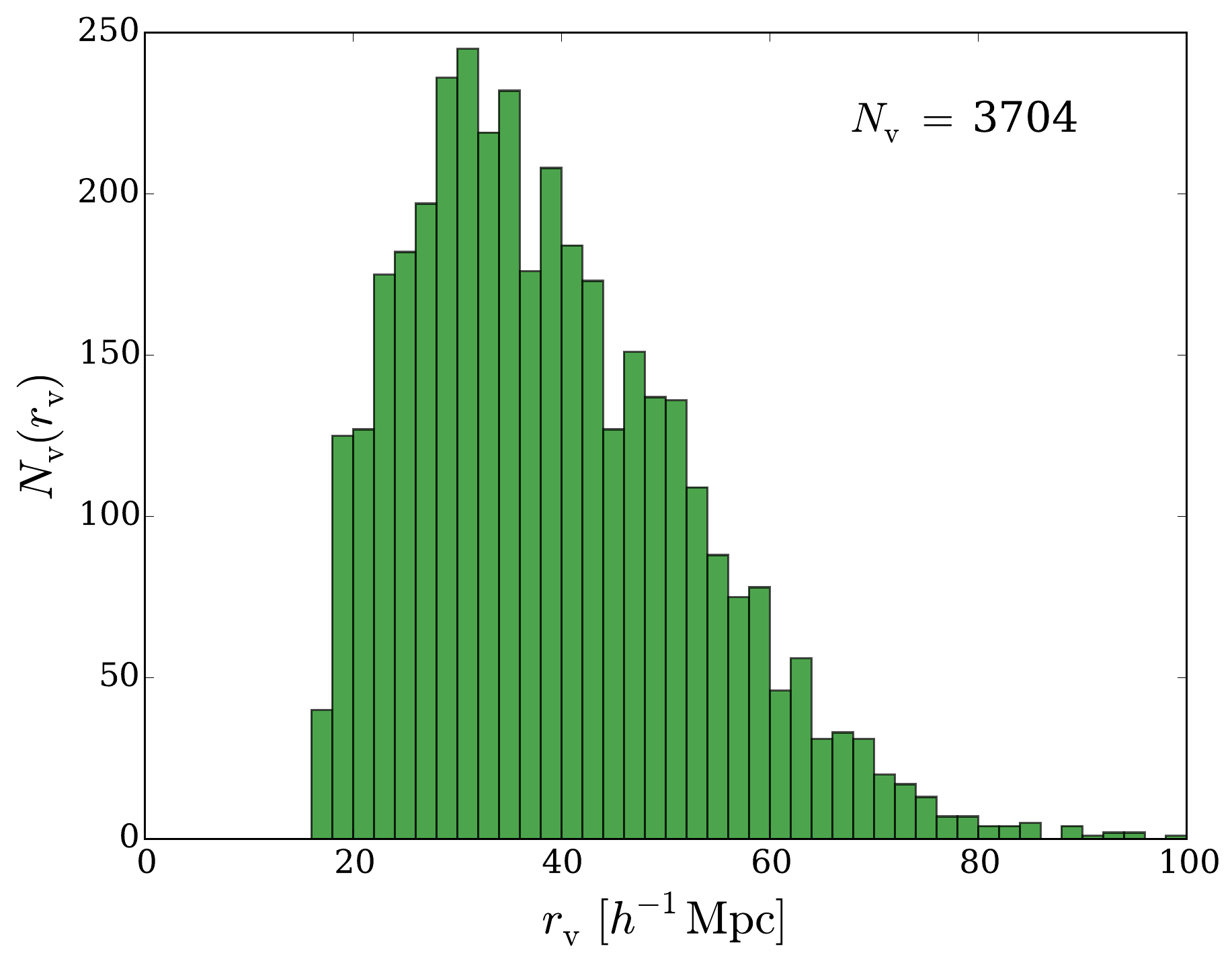}}
\caption{Void effective radius distribution in LOWZ (left) and CMASS (right).}
\label{Nv}
\end{figure*}

We choose to define void centers as volume-weighted barycenters of all the tracer particles making up each void. This can be achieved by calculating
\begin{equation}
 \mathbf{X} = \frac{\sum_j\mathbf{x}_jV_j}{\sum_jV_j}\;,
\end{equation}
where $\mathbf{x}_j$ is the coordinate vector of tracer $j$ with corresponding Voronoi-cell volume $V_j$. This center definition is very robust against Poisson fluctuations, as it is constrained by many tracer particles. Moreover, it preserves topological information about the void edges, which were used to define each void in the first place. The geometrical structure of watershed voids can be rather intricate and far from spherical. Yet it is convenient to assign an \emph{effective radius} $r_\void$ to each void, calculated as the radius of a sphere with the same total volume,
\begin{equation}
 r_\void = \left(\frac{3}{4\pi}\sum\nolimits_jV_j\right)^{1/3}\;.
\end{equation}
Figure~\ref{Nv} shows the distribution of effective void radii that we identified in the LOWZ and CMASS samples. In order to avoid contamination by spurious Poisson fluctuations, we discard voids with effective radii that are smaller than twice the mean galaxy separation at any given redshift, so we impose
\begin{equation}
 r_\void > 2\left[\frac{4\pi}{3}\bar{n}(z)\right]^{-1/3}\;,
\end{equation}
where $\bar{n}(z)$ is the mean galaxy density at redshift $z$. The final catalogs comprise $963$ voids with $15.7\hMpc\le r_\void\le78.8\hMpc$ in LOWZ, and $3704$ voids with $17.2\hMpc\le r_\void\le98.6\hMpc$ in CMASS. All of these voids are root nodes that do not overlap in volume. The BOSS tracer density is too low to probe a deeper void hierarchy with sub-voids.

\section{Analysis \label{sec:analysis}}

\subsection{Estimator}
The void-galaxy cross-correlation function $\xi^s(r,\mu)$ describes the average galaxy number-density fluctuation around voids in redshift space, $n(r,\mu)/\bar{n}-1$, which can be written as a sum over all $N_\gal$ galaxy positions $\mathbf{x}_j$ around each of the $N_\void$ voids with effective radius $r_{\void_i}$, located at $\mathbf{X}_i$~\cite{Hamaus2015}. Plugging this into equation~(\ref{xi_l}) to calculate the multipoles, we get
\begin{multline}
 \xi_\ell(r) = \int_0^1\left[\frac{1}{N_\void} \sum_i \frac{1}{\bar{n}_i} \sum_j \delta^\mathrm{D}\left(\frac{\mathbf{X}_i-\mathbf{x}_j+\mathbf{r}}{r_{\void_i}}\right)-1\right](1+2\ell)P_\ell(\mu)\mathrm{d}\mu = \\
 = \frac{1}{N_\void} \sum_i \frac{1}{\bar{n}_i} \sum_j(1+2\ell)P_\ell\left(\mu=\frac{\mathbf{r}_{ij}\cdot\mathbf{X}_i}{|\mathbf{r}_{ij}||\mathbf{X}_i|}\right) - \delta^\mathrm{K}_{0\ell} \;, \label{xi_le}
\end{multline}
where $\mathbf{r}_{ij}=-\mathbf{X}_i+\mathbf{x}_j$, $\delta^\mathrm{D}$ denotes a Dirac delta function and $\delta^\mathrm{K}$ a Kronecker delta. Thus, the multipoles can simply be calculated as weighted 1D-histograms of the distances between galaxies and void centers. The other option is to compute the 2D-histogram in $r$ and $\mu$ as a first step, and performing the $\mu$-integration thereafter. For sparsely distributed tracers, however, this can be numerically suboptimal, which is especially the case deep inside voids at small values of $\mu$~\cite{Cai2016}. Moreover, we normalize void-centric distances by the effective radius $r_{\void_i}$ of each void, such that the characteristic shape of the void profile is retained in the final correlation function. This procedure can also be referred to as \emph{void stacking}.

\begin{figure*}[!t]
\centering
\resizebox{\hsize}{!}{
\includegraphics{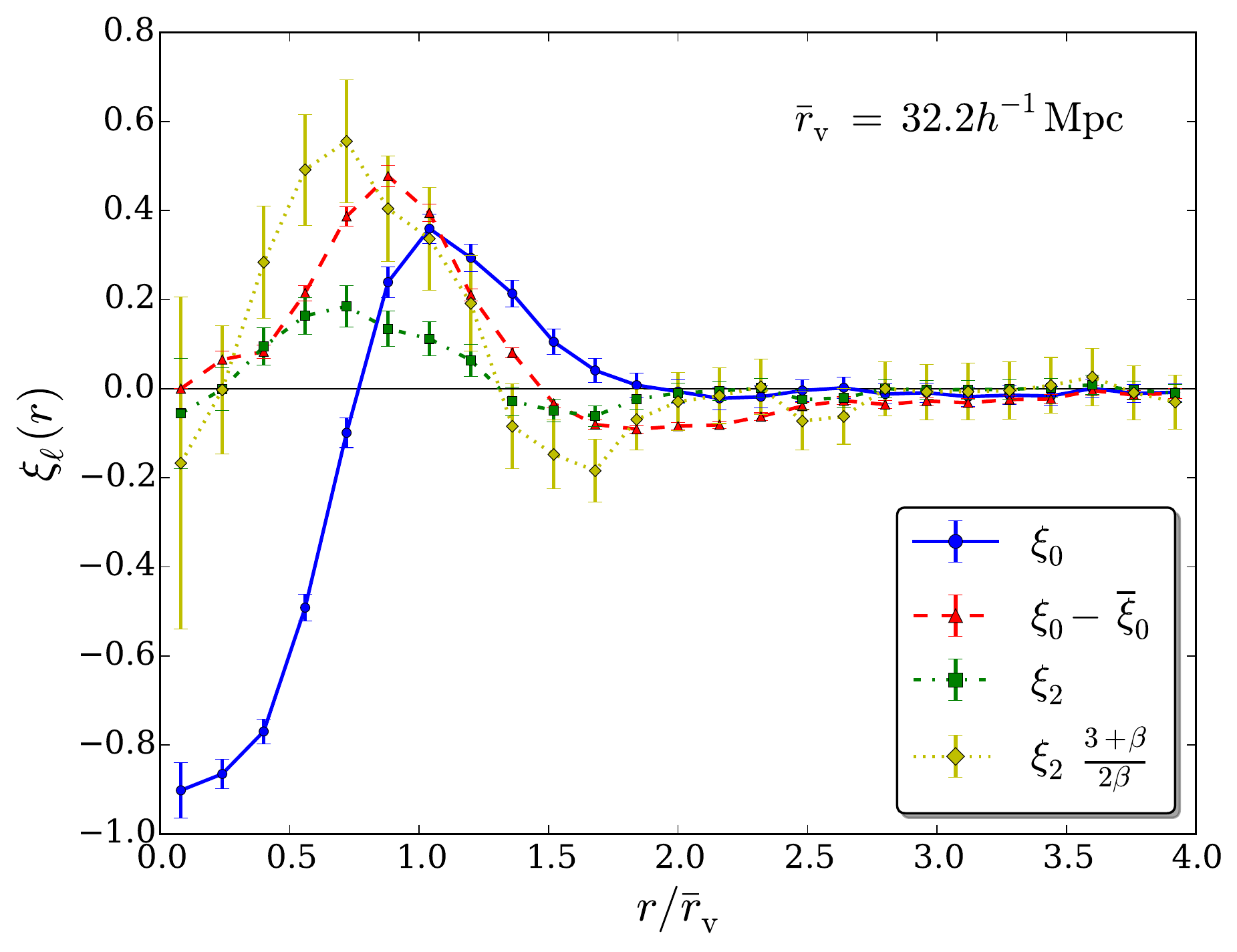}
\includegraphics{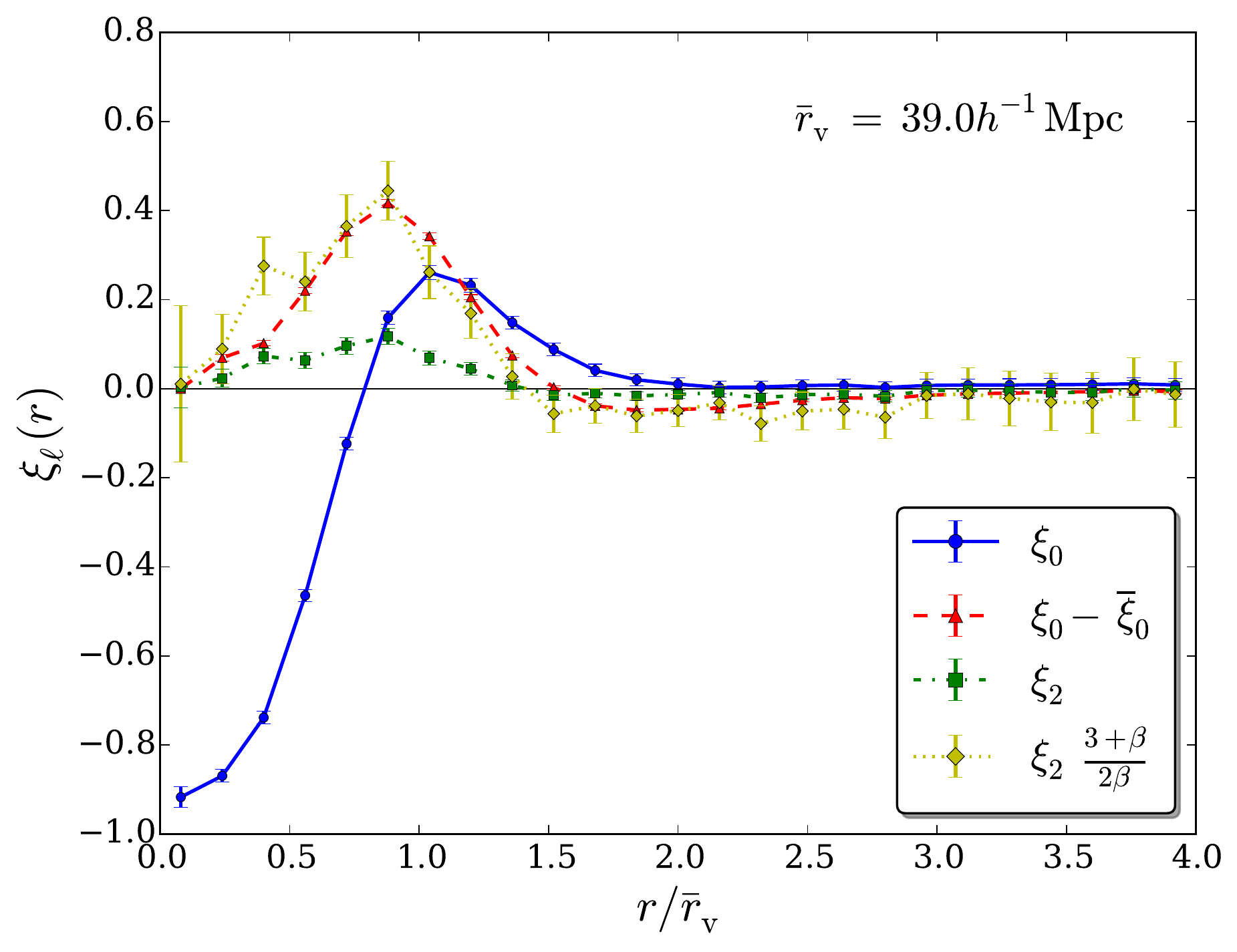}}
\caption{Multipoles of the void-galaxy cross-correlation function in LOWZ (left) and CMASS (right). In computing the ratio $(3+\beta)/2\beta$, we use the maximum-likelihood value obtained for $\beta$.} 
\label{Xvg}
\end{figure*}

\begin{figure*}[!t]
\centering
\resizebox{\hsize}{!}{
\includegraphics{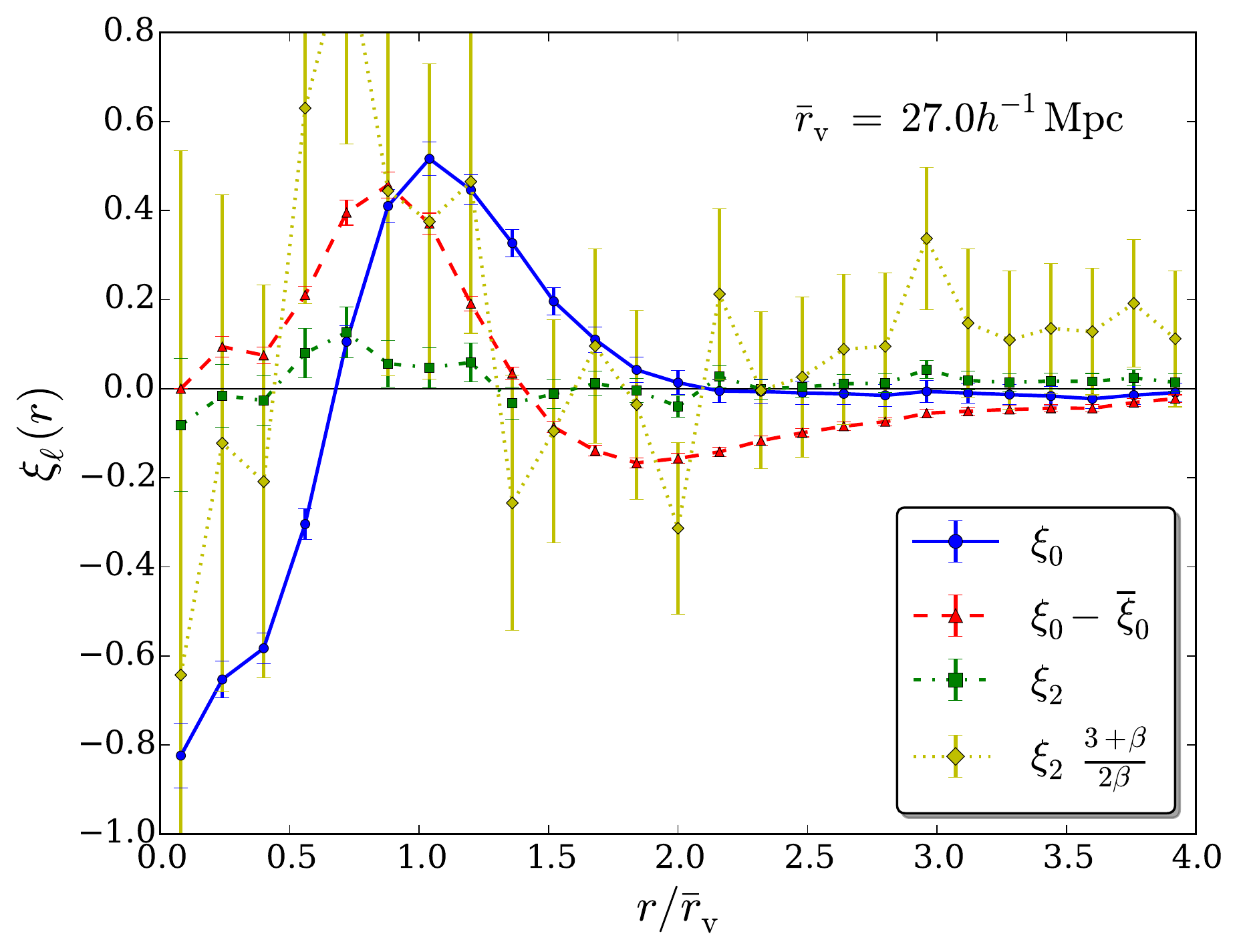}
\includegraphics{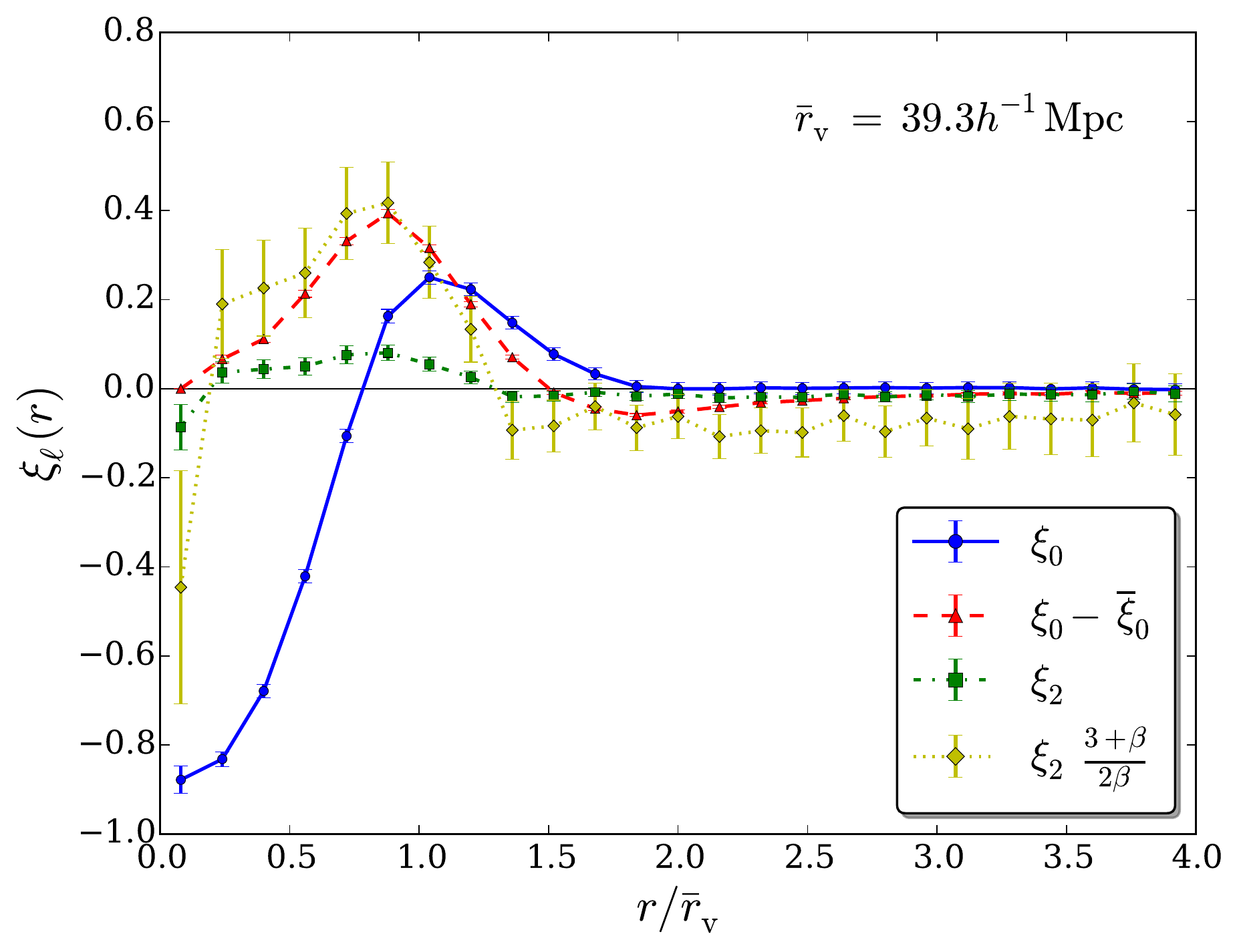}}
\caption{Same as figure~\ref{Xvg} for QPM mocks.}
\label{Xvg_mock}
\end{figure*}

Further complications in estimating the multipoles arise through survey boundary effects and a varying tracer number density with redshift. So instead of simply counting void-galaxy pairs as a function of their separation, it is conducive to additionally consider pair counts for randomly distributed objects of each type, exhibiting the same redshift distribution and survey boundaries. The \emph{Landy-Szalay} (LS) estimator combines all possible correlations between the data $D$ and the randoms $R$ to calculate the underlying correlation function of the data in a nearly optimal way~\cite{Landy1993}. For the void-galaxy cross-correlation function it reads~\cite{Achitouv2017}
\begin{equation}
 \xi^s(r,\mu) = \frac{\langle D_\void D_\gal \rangle - \langle D_\void R_\gal \rangle - \langle D_\gal R_\void \rangle + \langle R_\void R_\gal \rangle}{\langle R_\void R_\gal \rangle} \;, \label{LS}
\end{equation}
where angled brackets denote normalized pair counts at separation $r$ and $\mu$. In order to compute this estimator as a function of distance normalized to the mean effective void radius $\bar{r}_\void$, we simply divide every void-centric separation by its corresponding $r_{\void_i}$. Since the random void positions $R_\void$ have no effective radii associated to them, we can choose the mean effective radius $\bar{r}_\void$ of the data voids for each rescaling. The multipoles of the LS-estimator can be obtained via applying equation~(\ref{xi_l}), which implies that all the correlations in equation~(\ref{LS}) have to be computed in bins of $r$ and $\mu$ before taking the ratio. However, we have investigated the $\mu$-dependence of $\langle R_\void R_\gal \rangle$ by calculating its quadrupole, finding it to be consistent with zero on the scales relevant to our analysis. Therefore, we can pull it out of the $\mu$-integral and simply compute the one-dimensional weighted histograms via equation~(\ref{xi_le}) for each correlator in equation~(\ref{LS}). The monopole of $\langle R_\void R_\gal \rangle$ varies on the sub-percent level, but as it appears on both sides of equation~(\ref{xi_0_2}), it does not play a role in the determination of $\beta$ anyway. Furthermore, we have also found the monopole and quadrupole of the combination $-\langle D_\gal R_\void \rangle + \langle R_\void R_\gal \rangle$ to be negligible for our purposes. Thus, for estimating the multipoles of the void-galaxy cross-correlation function, we will simply use
\begin{equation}
 \xi_\ell(r) \simeq \langle D_\void D_\gal \rangle_\ell - \langle D_\void R_\gal \rangle_\ell \;. \label{LS_l}
\end{equation}
Figure~\ref{Xvg} depicts the multipoles $\xi_0(r)$ and $\xi_2(r)$ out to a separation of four times the mean effective radius of the entire void sample in both LOWZ and CMASS. We also computed the hexadecapole $\xi_4(r)$, and verified that it is consistent with zero everywhere, in agreement with equation~(\ref{xi_s}). The monopole shows the characteristic shape of the void-density profile, with a deeply under-dense core and an over-dense ridge at $r=\bar{r}_\void$~\cite{Hamaus2014b}. Also the quadrupole is detected with high significance, exhibiting positive values inside voids, but changing sign slightly outside of their mean effective radius, when $\xi(r)=\xibar(r)$ at $r\simeq1.5\bar{r}_\void$. This behaviour is in agreement with simulation results of the radial velocity profile of voids, showing a transition between outflow and infall of tracer particles at similar scales~\cite{Hamaus2014b}. The functional form of the quadrupole nicely agrees with the combination $\xi_0-\xibar_0$, as predicted by equation~(\ref{xi_0_2}). We notice a fairly high amplitude of the quadrupole from the LOWZ sample, which already suggests a higher value for $\beta$ than what is expected from the CMASS sample, but we will quantify this in the following sections. Figure~\ref{Xvg_mock} presents the multipoles from the QPM mock catalog. The similarity with the real data is striking, except maybe for the quadrupole amplitude in the LOWZ sample, which is lower in the mocks. When multiplying the quadrupole with our best-fit value for the ratio $(3+\beta)/2\beta$, this results in larger error bars than in the real data, because the inferred value of $\beta$ is smaller in the mocks.

\subsection{Likelihood}
\begin{figure*}[!t]
\centering
\resizebox{\hsize}{!}{
\includegraphics{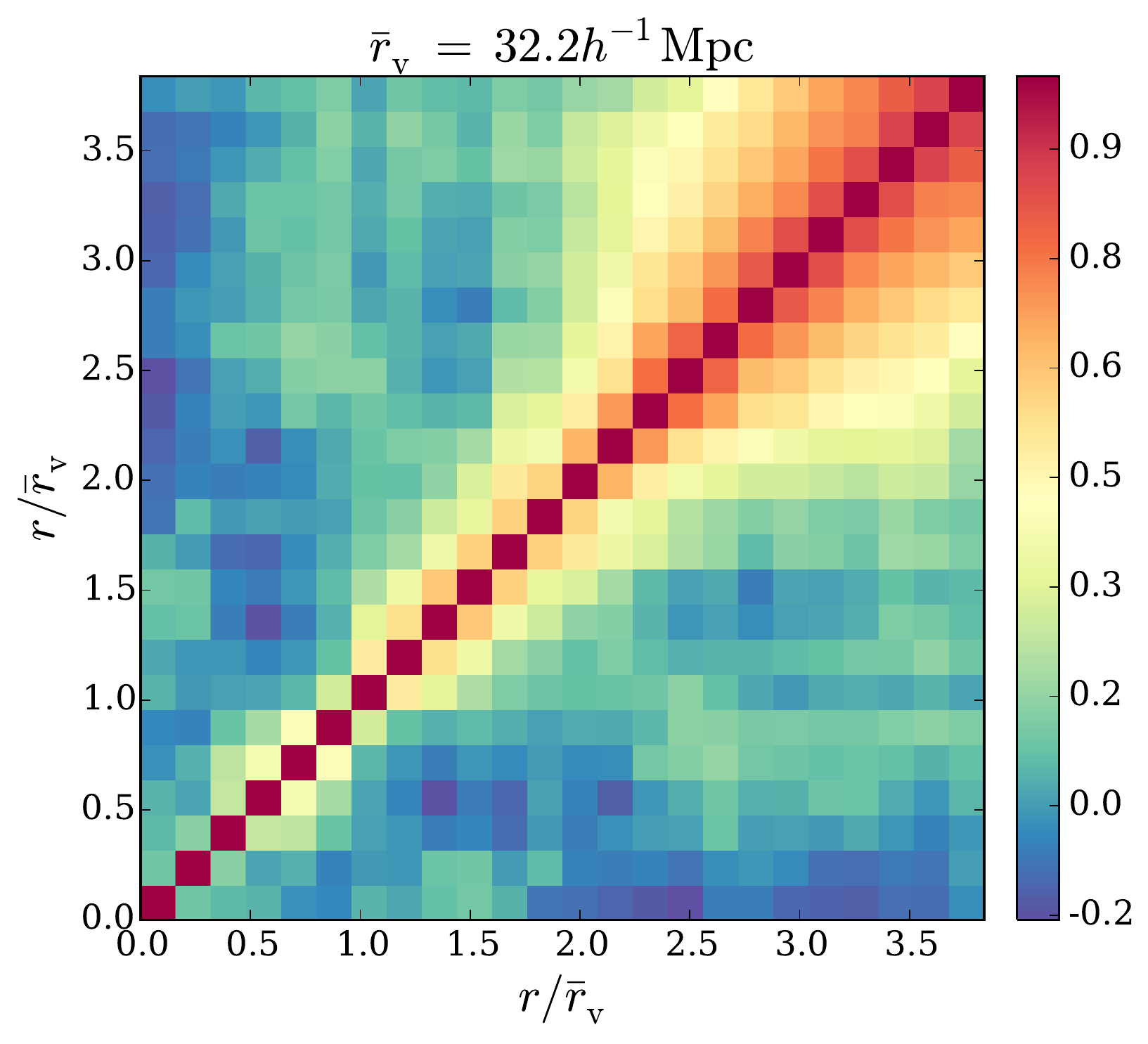}
\includegraphics{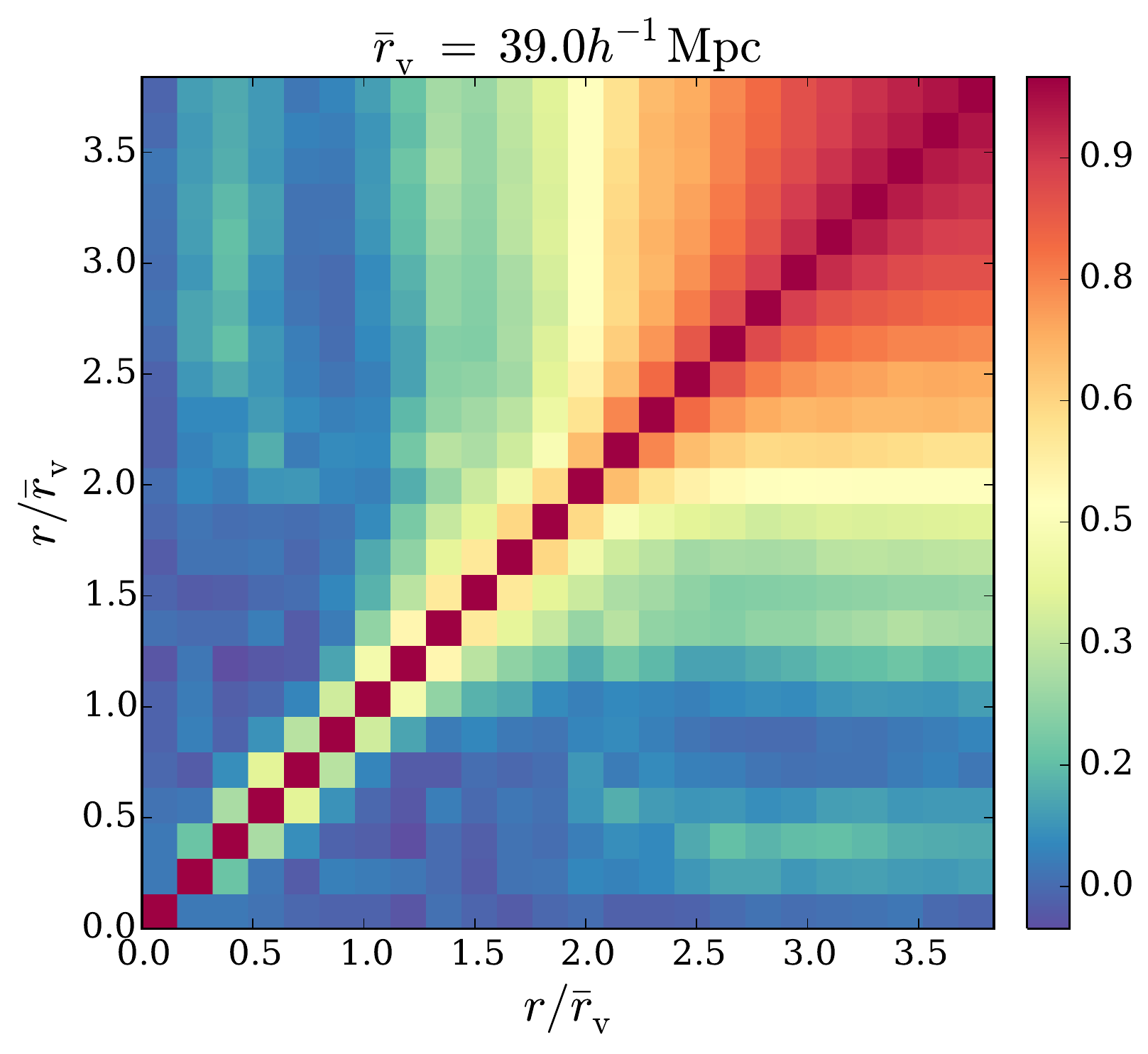}}
\caption{Correlation matrix for residuals between multipoles in LOWZ (left) and CMASS (right).}
\label{Cov}
\end{figure*}

In order to determine the growth rate from measurements of the monopole and quadrupole of the void-galaxy cross-correlation function, equation~(\ref{xi_0_2}) can be directly solved for $\beta$. However, this entails taking ratios of low signal-to-noise data, so we follow the approach of reference~\cite{Cai2016} and fit for $\beta$ in $N$ radial bins $r_i$, assuming a Gaussian likelihood
\begin{equation}
 L(\xi_\ell|\beta) = \frac{1}{(2\pi)^{N/2}\sqrt{\det\C}}\exp\left(-\frac{1}{2}\sum\limits_{i,j}^N\varepsilon_i\,\C_{ij}^{-1}\varepsilon_j\right)\;, \label{likelihood}
\end{equation}
with residuals $\varepsilon_i =  \xi_2(r_i) - \frac{2\beta}{3+\beta}\left[\xi_0(r_i) - \xibar_0(r_i)\right]$ and covariance matrix $\C_{ij} = \langle\varepsilon_i\varepsilon_j\rangle$. Note that $\C$ explicitly depends on $\beta$, so the normalization of the likelihood with $\sqrt{\det\C}$ must be taken into account in its maximization procedure. We estimate the covariance matrix via jackknife resampling the multipoles of the void-galaxy cross-correlation function. This can be achieved $N_\void$ times by removal of each individual void one by one in evaluating equation~(\ref{LS_l}). The covariance is then estimated as
\begin{equation}
 \C_{ij} = \frac{N_\void-1}{N_\void}\sum_{k=1}^{N_\void}\left(\varepsilon_i^{\!(k)}-\tilde{\varepsilon}_i\right)\left(\varepsilon_j^{\!(k)}-\tilde{\varepsilon}_j\right)\;, \label{cov}
\end{equation}
where $\varepsilon_i^{\!(k)}$ is the residual at $r_i$ after discarding void $k$, and $\tilde{\varepsilon}_i = N_\void^{-1}\sum_k\varepsilon_i^{\!(k)}$ is the mean residual over all $N_\void$ jackknife samples. We follow the same procedure to compute error bars on the individual multipoles, shown as the square root of the diagonal elements in their covariance matrix in figures~\ref{Xvg} and~\ref{Xvg_mock}. The covariance matrix of the residuals after normalization by its diagonal (correlation matrix) is shown in figure~\ref{Cov} for both LOWZ and CMASS, using the corresponding maximum-likelihood value for $\beta$ in its estimation. We notice an onset of significant covariance only outside the mean effective void radius, whereas below $\bar{r}_\void$ the residuals tend to be uncorrelated, or even weakly anti-correlated.

\begin{figure*}[!t]
\centering
\resizebox{\hsize}{!}{
\includegraphics{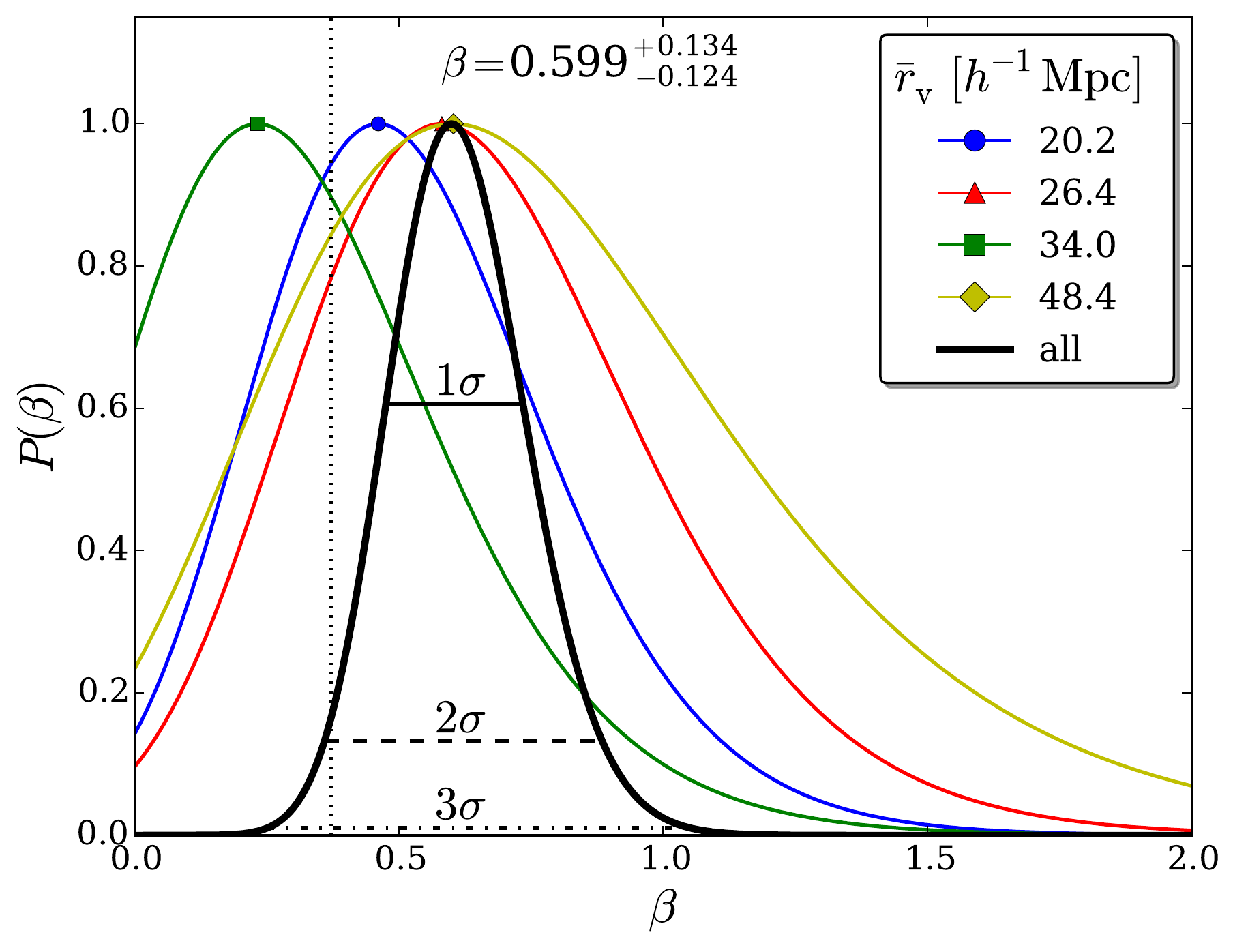}
\includegraphics{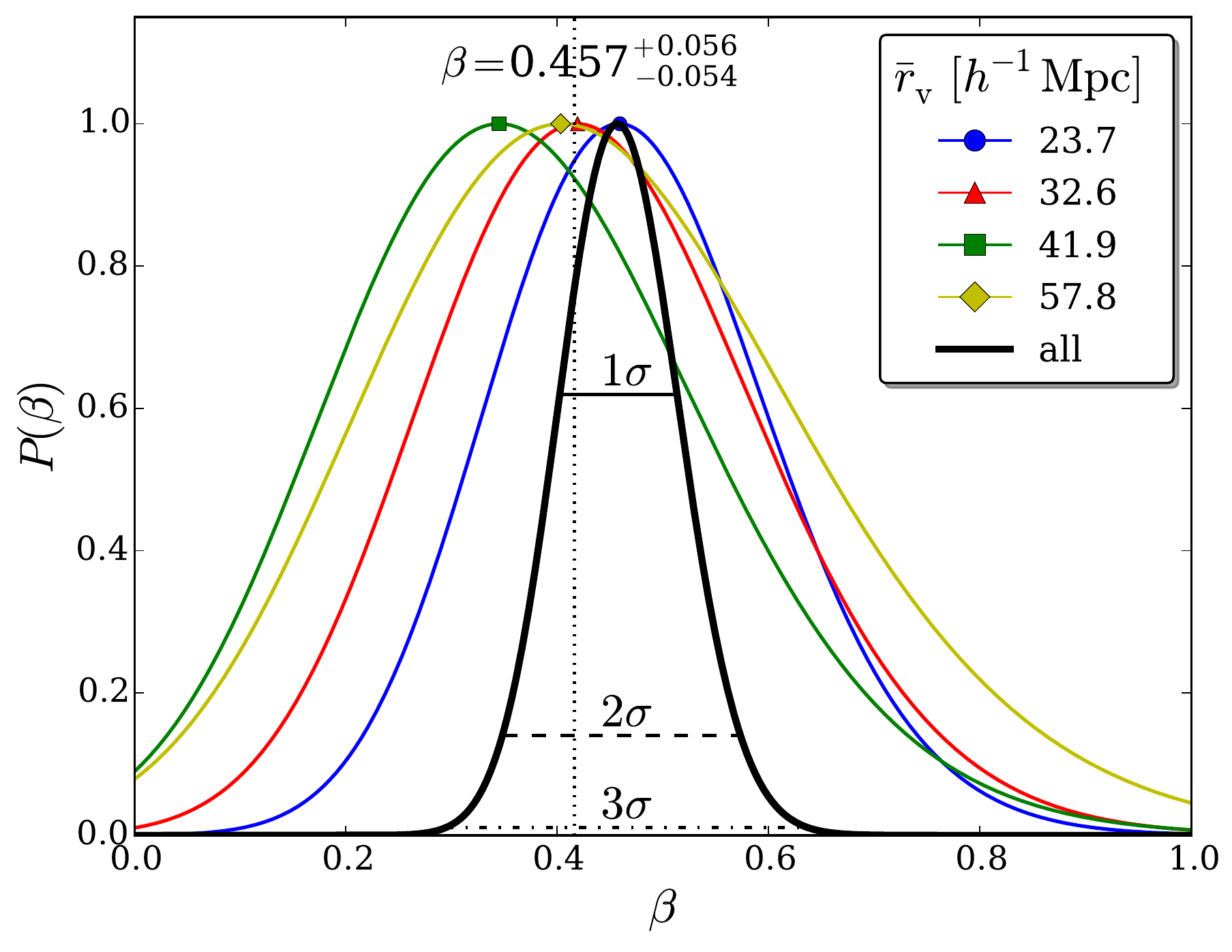}}
\caption{Posterior distribution for $\beta$ in LOWZ (left) and CMASS (right) from four bins in void radius (colored with symbols), and for all voids (bold black). The dotted lines indicate $\beta=\Omega_\mathrm{m}^{\;\gamma}(\bar{z})/b$.}
\label{beta}
\end{figure*}

\begin{figure*}[!t]
\centering
\resizebox{\hsize}{!}{
\includegraphics{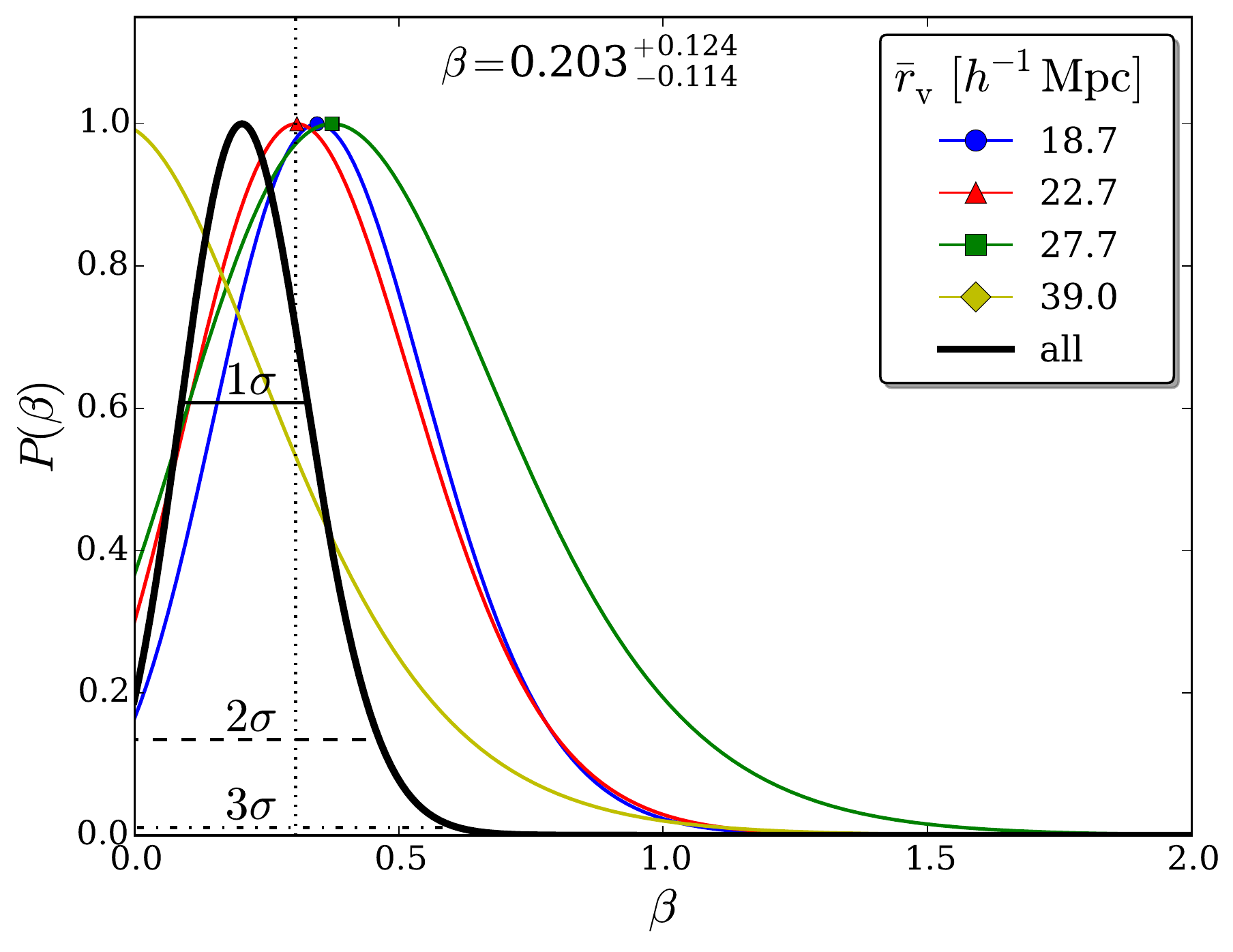}
\includegraphics{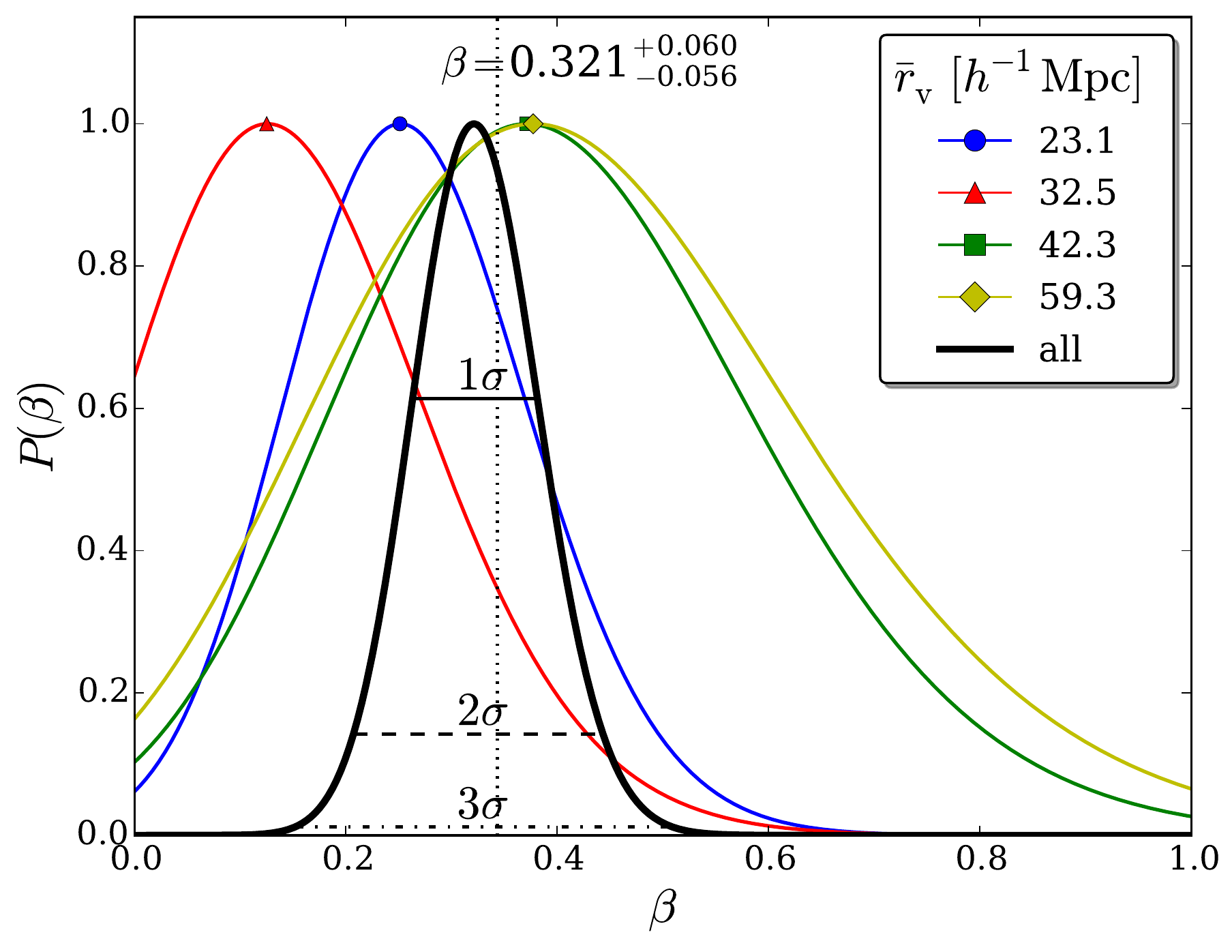}}
\caption{Same as figure~\ref{beta} for QPM mocks.}
\label{beta_mock}
\end{figure*}

We numerically minimize the negative logarithm of the likelihood in equation~(\ref{likelihood}) with respect to $\beta$ and find its most likely value including the $1\sigma$, $2\sigma$, and $3\sigma$ confidence intervals. The value of $\beta$ is not restricted to be positive, it may take on any value in the inference process. Complete posterior probability distribution functions for $\beta$ in LOWZ and CMASS are presented in figure~\ref{beta}, both for the full void sample, as well as for four bins in void radius as a consistency check. The bins roughly contain the same number of voids, such that they carry the same statistical weight. The posteriors from the bins are all consistent with each other and with the posterior of the full sample, we do not observe any significant trend with void size. Moreover, the CMASS sample agrees to within $1\sigma$ with the expectation from GR and our fiducial $\Lambda$CDM cosmology, $\beta\simeq0.416$ (vertical dotted line). However, the inferred probability distribution for $\beta$ in the LOWZ sample is marginally consistent with the expected value of $\beta\simeq0.374$ to within $2\sigma$. This quantifies the high value of the quadrupole as seen in figure~\ref{Xvg}, but does not constitute a significant enough tension to look out for explanations of other than statistical origin. Yet, we do find the QPM mocks to be consistent with their input cosmology to within $1\sigma$, as evident from figure~\ref{beta_mock}. An overview of our constraints on $\beta$ is given in table~\ref{table}.

\section{Discussion\label{sec:discussion}}
We have investigated a wide range of systematic effects that could potentially impact our results. We tested our estimator for the void-galaxy cross-correlation function extensively. First of all, we have made sure that equation~(\ref{LS_l}) is a valid approximation for equation~(\ref{LS}), by comparing the resulting multipoles and final $\beta$ constraints. We also applied weights to each galaxy in the pair counting, which account for various systematics, such as fiber collisions, redshift failures, stellar contamination, and seeing~\cite{Reid2016}. However, we observe no noticeable differences with or without these weights. A similar conclusion was reached for applying FKP weights~\cite{Feldman1994}, so we refrained from using any weighting schemes in our analysis. Further tests on the covariance matrix estimation have been performed, where we exchanged jackknife resampling with bootstrapping on both void and galaxy samples. The resulting error bars remain compatible among the different techniques. Ignoring the $\beta$-dependence of the covariance matrix in equation~(\ref{cov}) affects the maximum-likelihood values and their confidence intervals on the level of $10\%$. A removal of voids that are defined by less than $20$ galaxies also did not affect any of our results, suggesting our analysis to be insensitive to a potential contamination of spurious Poisson voids.

\begin{figure*}[!t]
\centering
\resizebox{\hsize}{!}{
\includegraphics{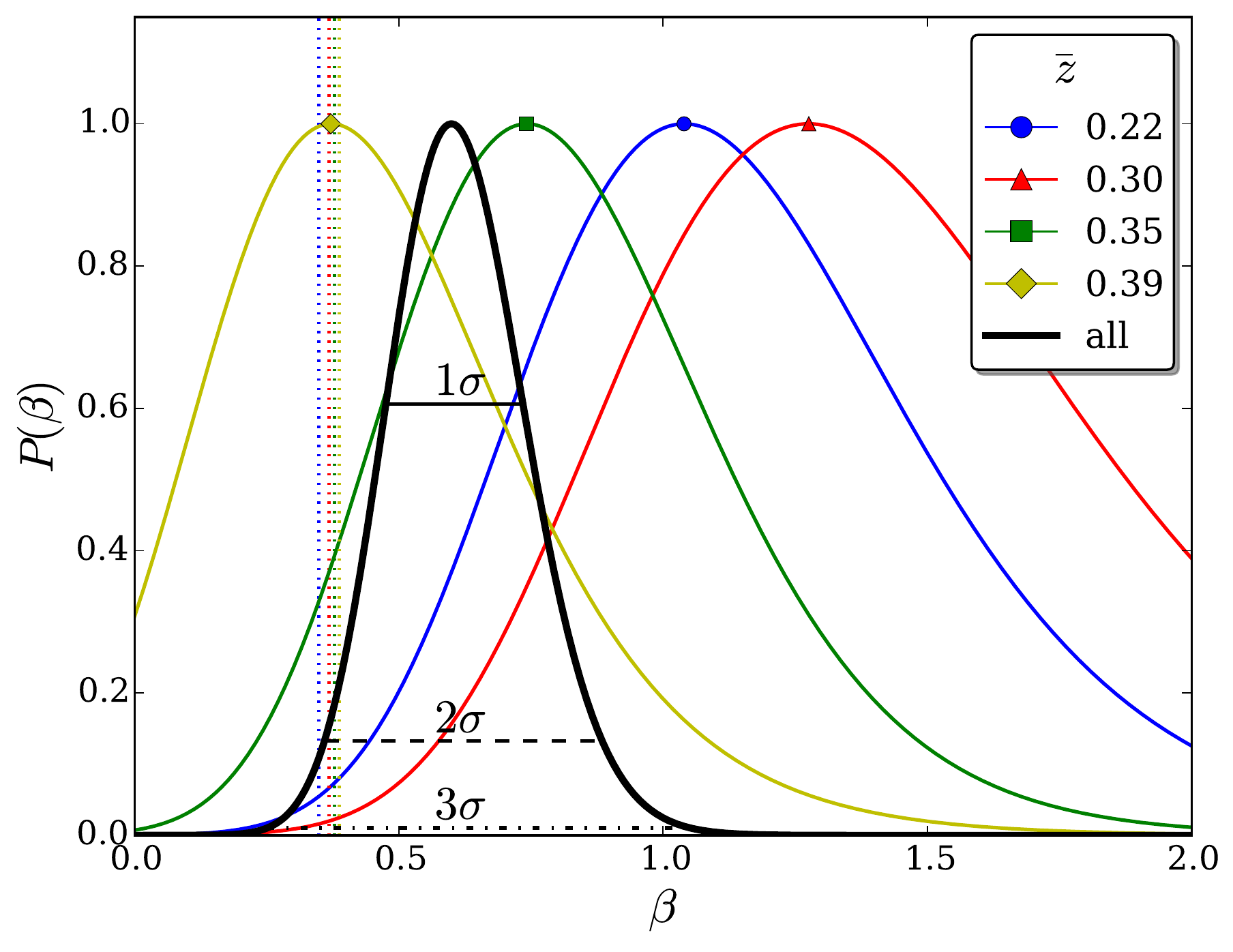}
\includegraphics{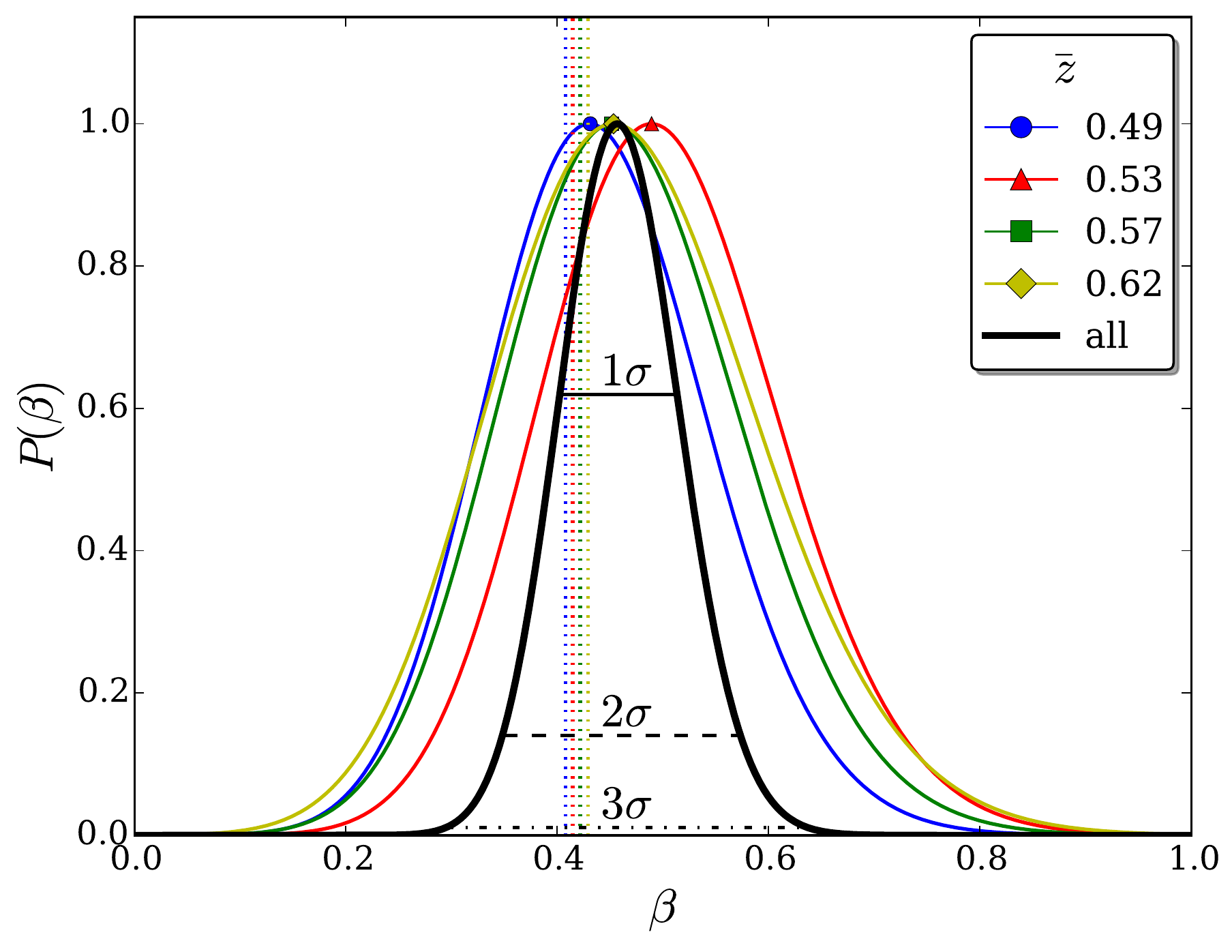}}
\caption{Posterior distribution for $\beta$ in LOWZ (left) and CMASS (right) from four bins in void redshift (colored with symbols), and for all voids (bold black). The dotted lines indicate $\beta=\Omega_\mathrm{m}^{\;\gamma}(\bar{z})/b$.}
\label{betaz}
\end{figure*}

\begin{figure*}[!t]
\centering
\resizebox{\hsize}{!}{
\includegraphics{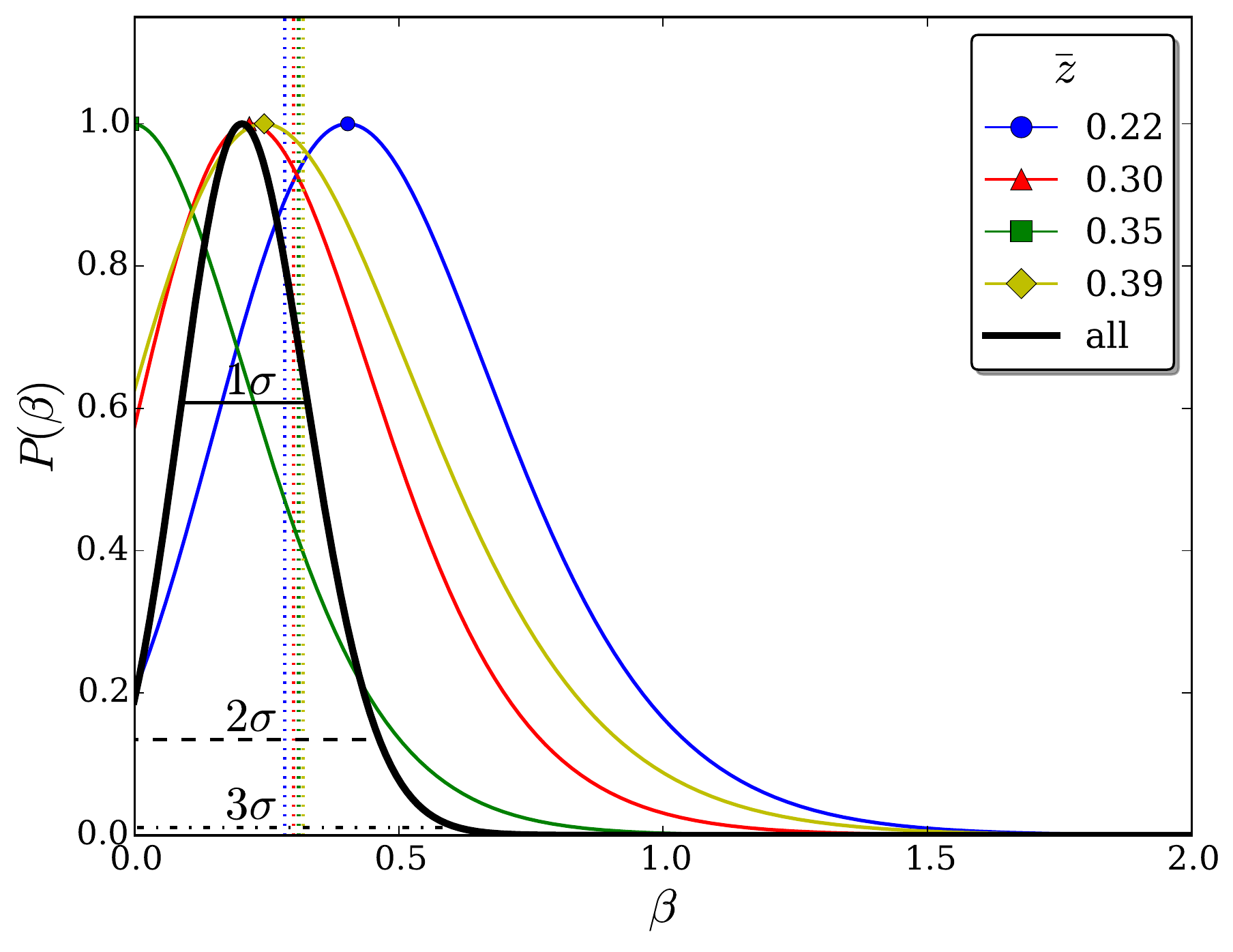}
\includegraphics{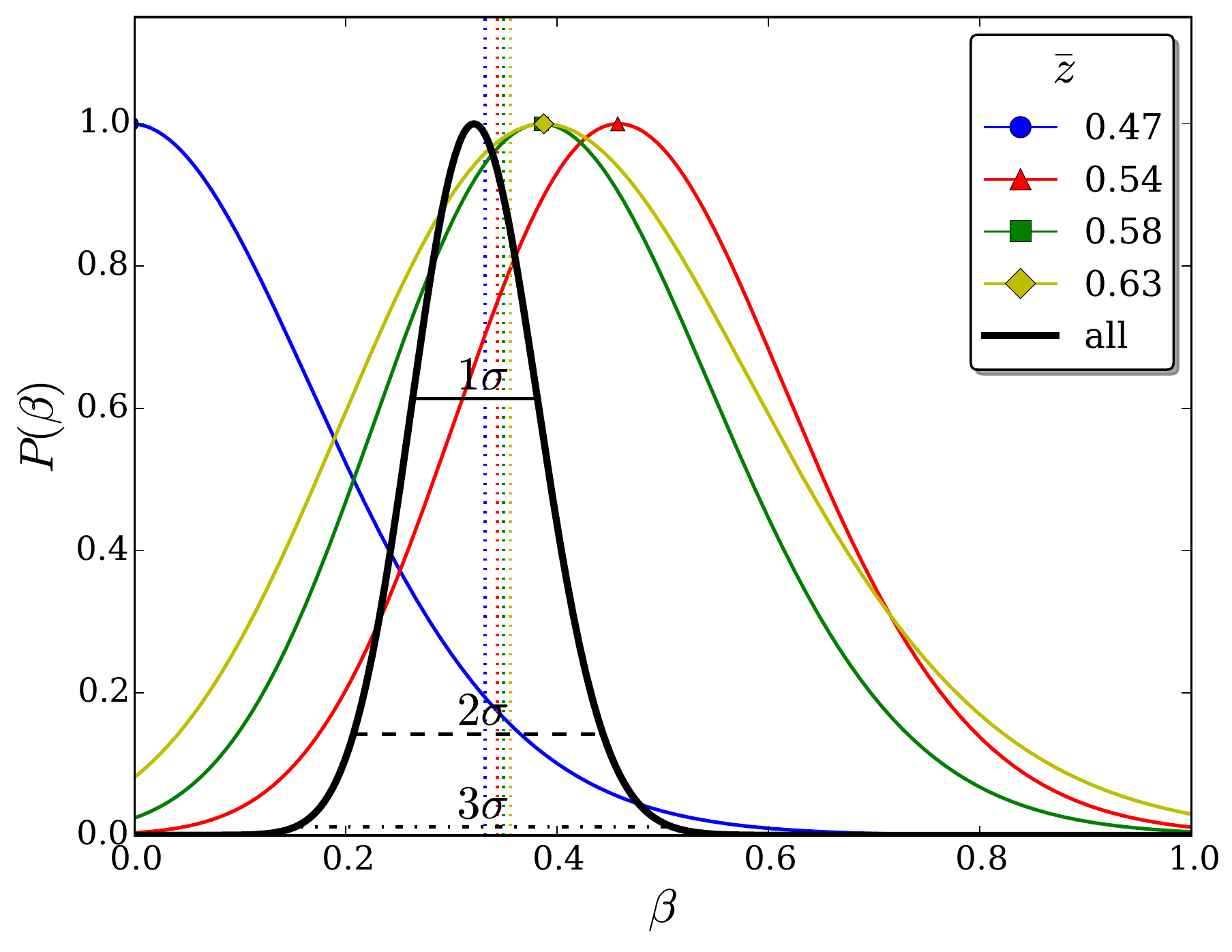}}
\caption{Same as figure~\ref{betaz} for QPM mocks.}
\label{betaz_mock}
\end{figure*}

\begin{figure*}[!p]
\centering
\resizebox{\hsize}{!}{
\includegraphics[width=\hsize,clip]{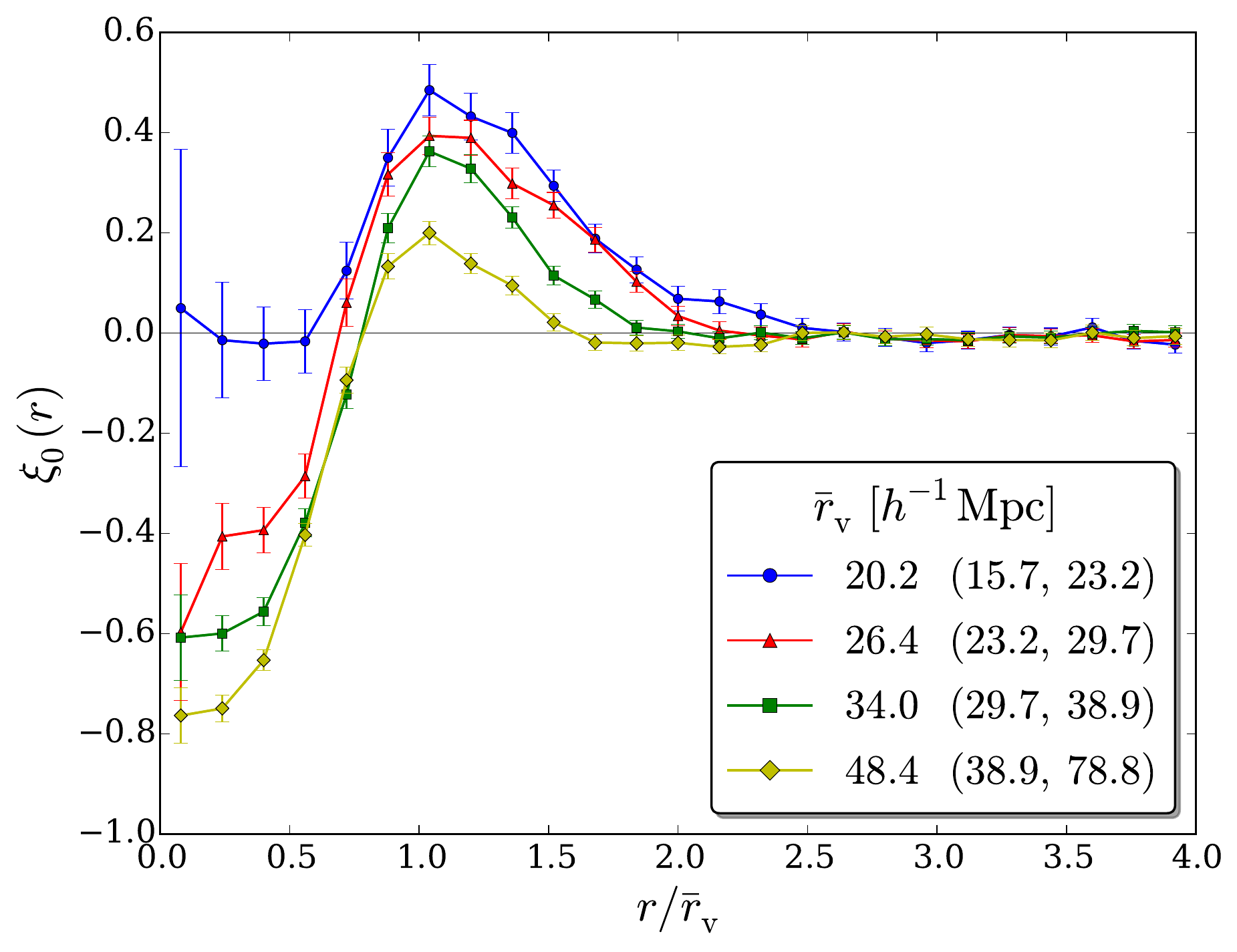}
\includegraphics[width=\hsize,clip]{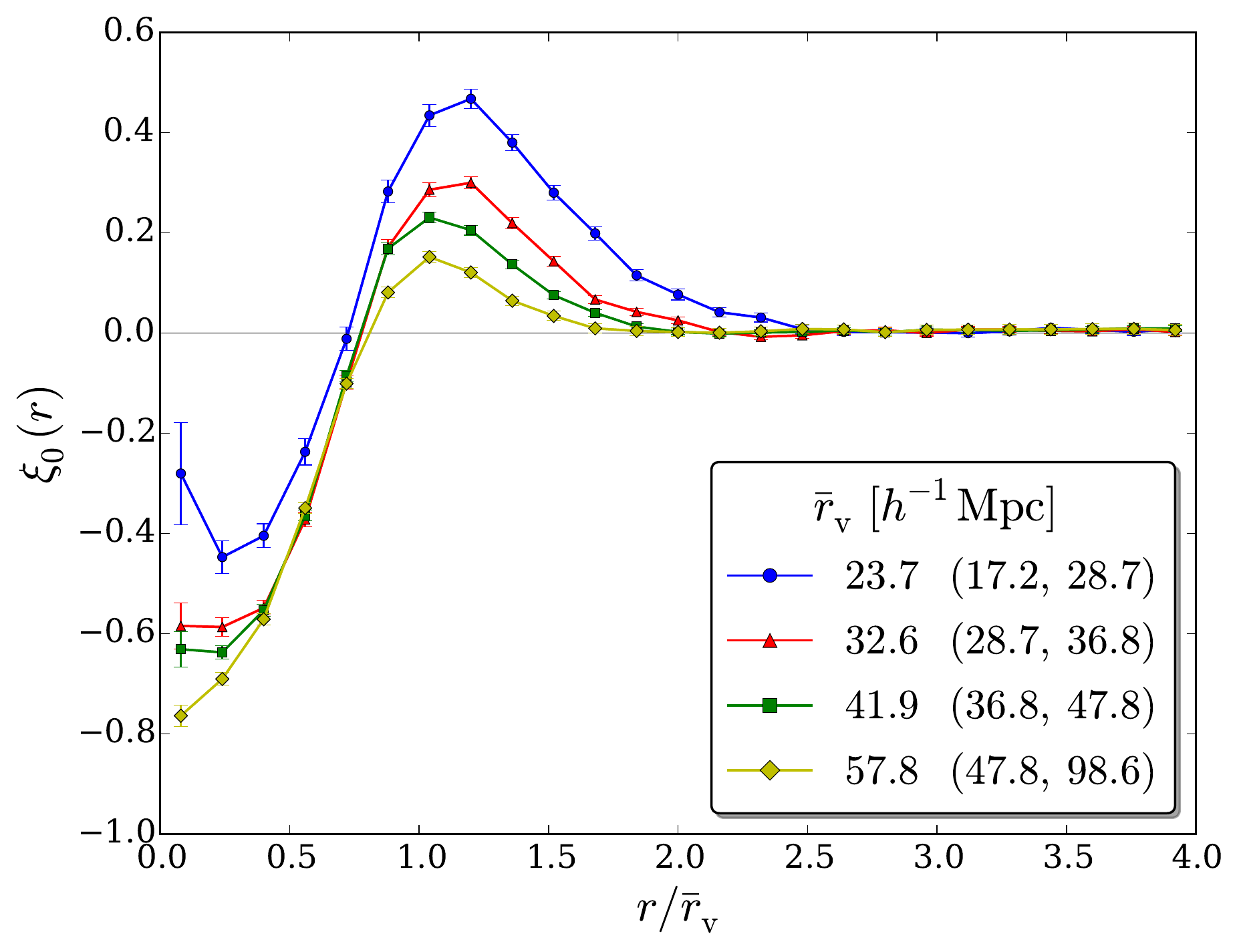}}
\resizebox{\hsize}{!}{
\includegraphics[width=\hsize,clip]{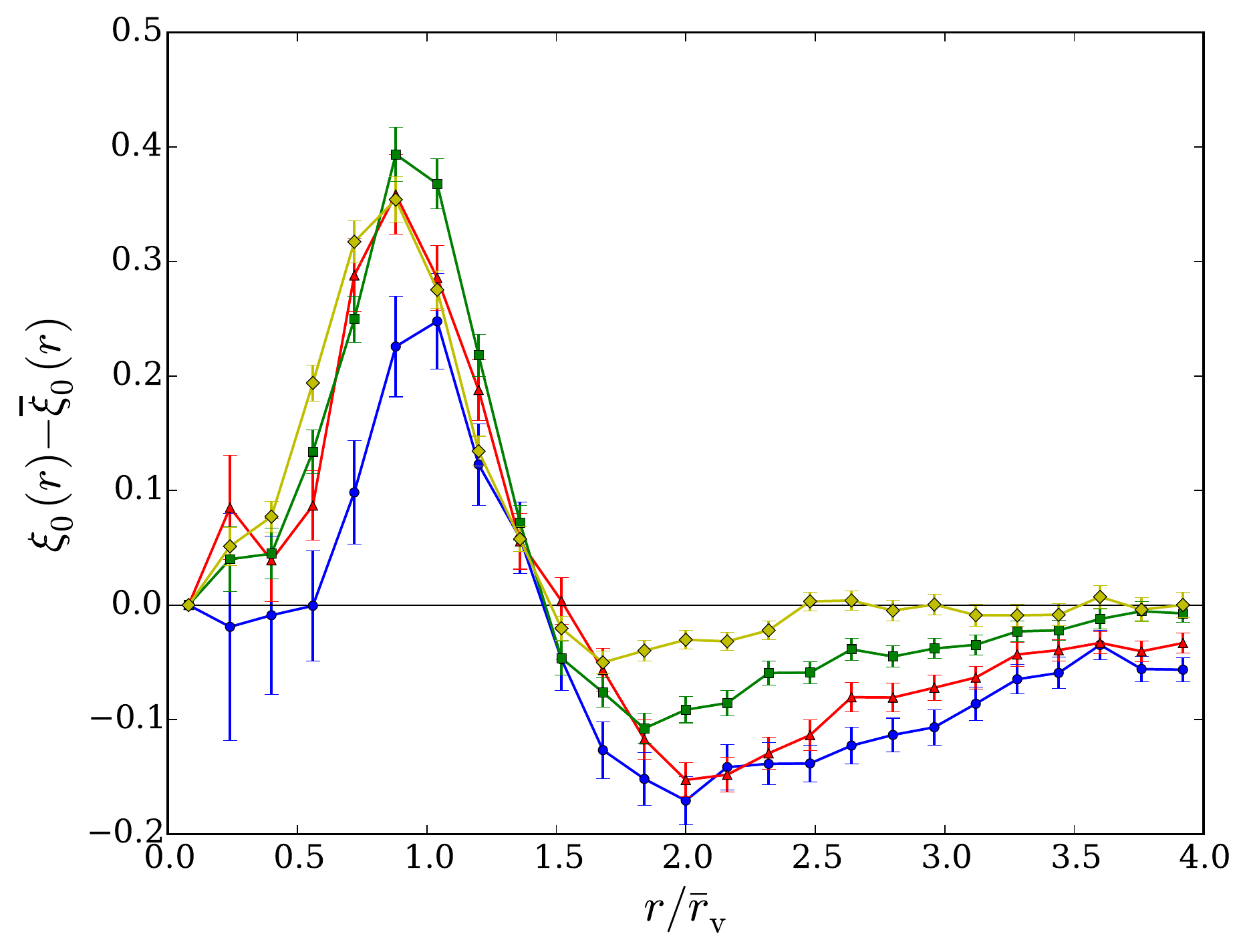}
\includegraphics[width=\hsize,clip]{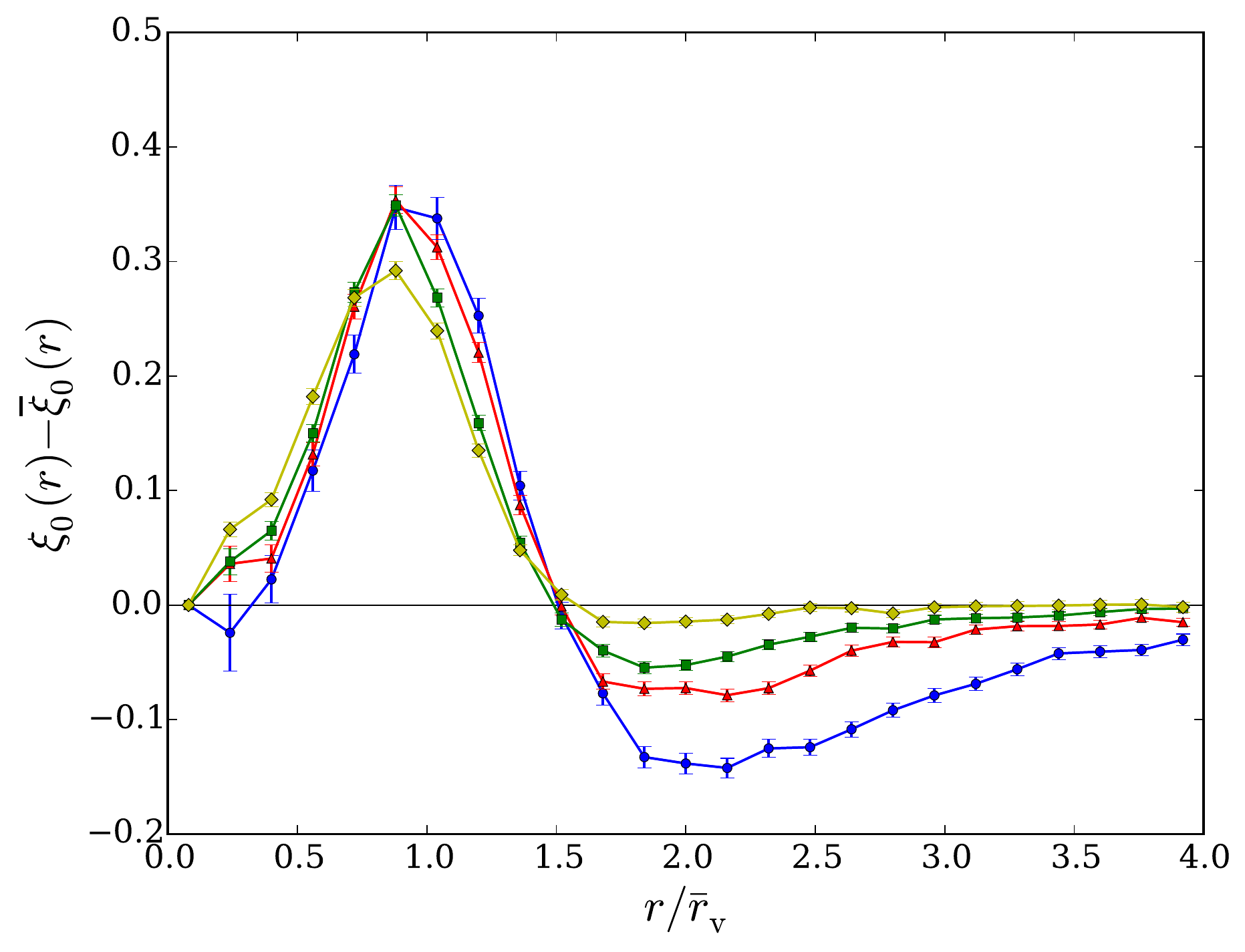}}
\resizebox{\hsize}{!}{
\includegraphics[width=\hsize,clip]{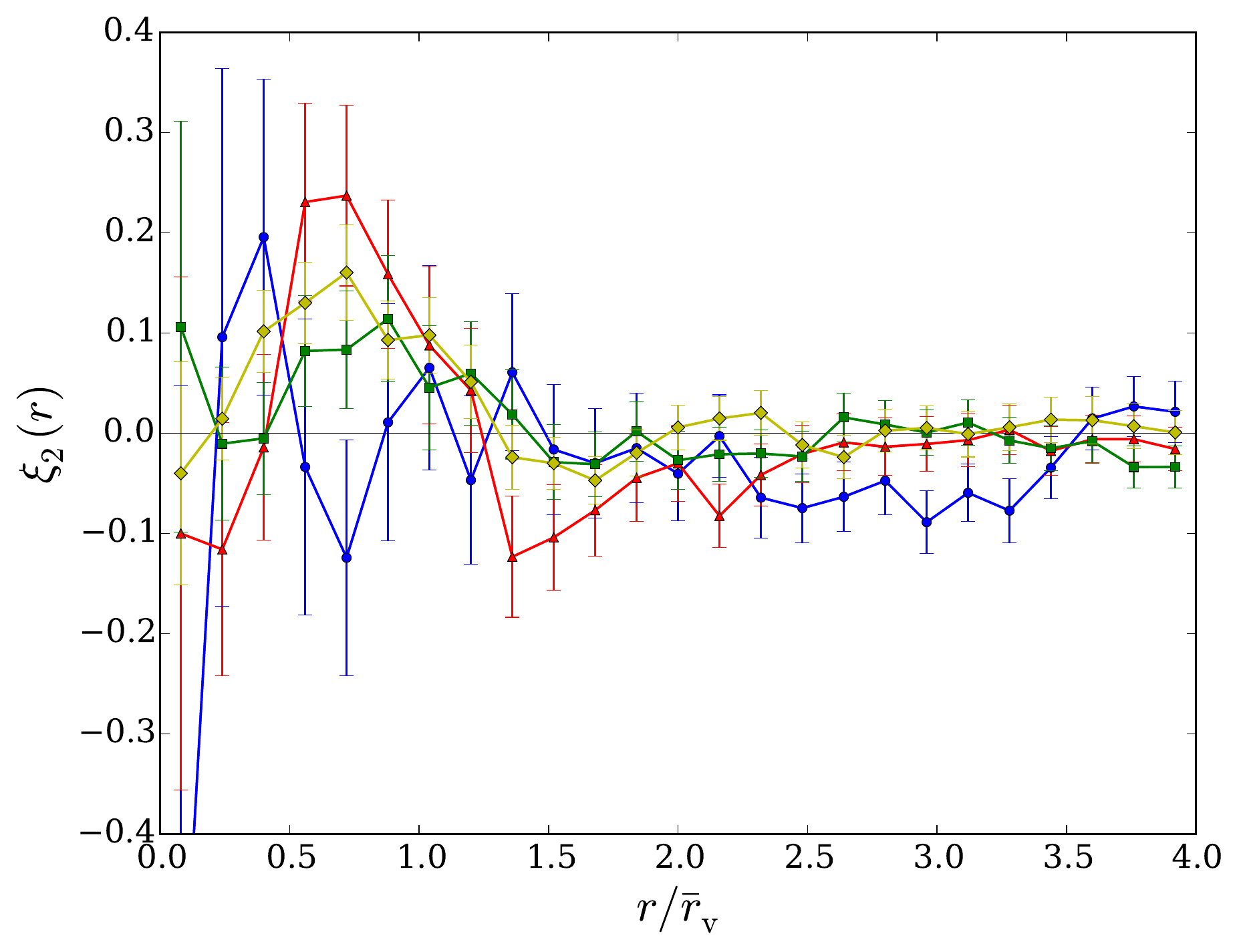}
\includegraphics[width=\hsize,clip]{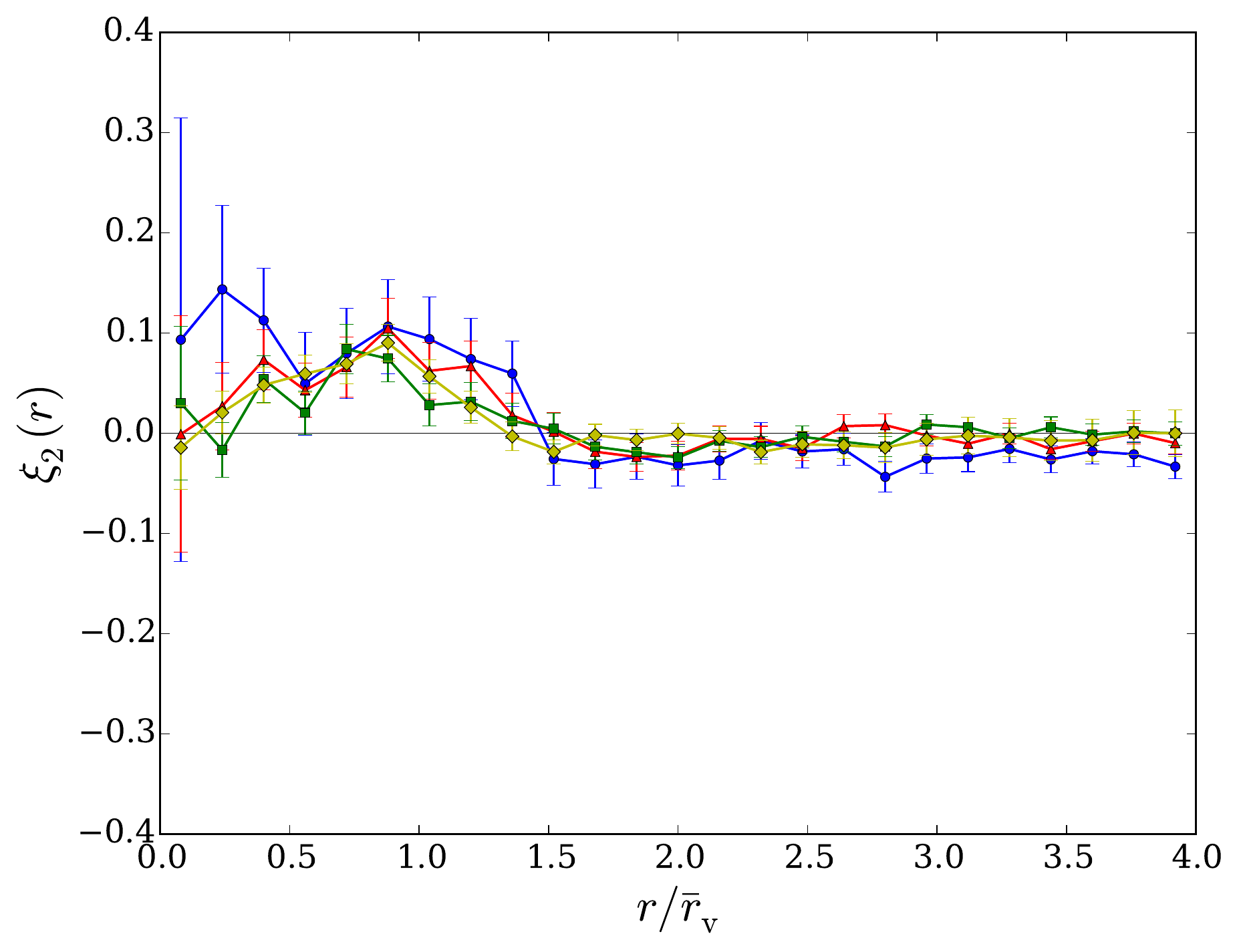}}
\caption{Multipoles of the void-galaxy cross-correlation function in LOWZ (left) and CMASS (right) for four bins in void radius. The bin edges are given in parenthesis in the figure legend.}
\label{Xvgi}
\end{figure*}

Although we already discussed the independence of our final constraints on void size in the previous section, for completeness we show the multipoles extracted from the same four bins in void effective radius in figure~\ref{Xvgi}. The monopole clearly evolves with void size, as the over-dense ridge feature becomes more prominent for smaller voids. A similar evolution is discernible for the quadrupole, albeit with larger noise: it follows the same trend as the combination $\xi_0-\xibar_0$, suggesting higher infall velocities in the outskirts of smaller voids. All of these observations are in excellent agreement with previous simulation results~\cite{Hamaus2014b}. Furthermore, the correspondence between the multipoles in their dependence on $r_\void$ is consistent with equation~(\ref{xi_0_2}) in each case, corroborating the validity of this equation. Only the smallest bin in void radius could potentially raise a cause of concern. Here the innermost radial bins of the multipoles are most affected by the sparsity of tracers, as the bin values approach the mean galaxy separation and hence suffer from poorer statistics~\cite{Sutter2014a}. Especially the LOWZ sample exhibits a fairly high monopole at small void-centric distances for the smallest bin in void radius. However, this bin is not driving the high value for $\beta$, as is evident from figure~\ref{beta}, so even discarding it would not change our conclusions.

\begin{table}[h]
\centering
\begin{tabular}{c|cccc}
 LOWZ & $\beta$ & $1\sigma$ & $2\sigma$ & $3\sigma$ \\
\hline\hline\\[-9pt]
 $\bar{r}_\void=20.2\hMpc$ & $0.461$ $(0.361)$ & $^{+0.298}_{-0.254}$ & $^{+0.664}_{-0.476}$ & $^{+1.119}_{-0.672}$ \\[3pt]
 $\bar{r}_\void=26.4\hMpc$ & $0.581$ $(0.374)$ & $^{+0.356}_{-0.298}$ & $^{+0.796}_{-0.552}$ & $^{+1.281}_{-0.748}$ \\[3pt]
 $\bar{r}_\void=34.0\hMpc$ & $0.233$ $(0.375)$ & $^{+0.324}_{-0.270}$ & $^{+0.728}_{-0.504}$ & $^{+1.245}_{-0.708}$ \\[3pt]
 $\bar{r}_\void=48.4\hMpc$ & $0.603$ $(0.378)$ & $^{+0.498}_{-0.386}$ & $^{+1.093}_{-0.668}$ & $^{+1.397}_{-0.844}$ \\[6pt]
 $\bar{z}=0.22$ & $1.040$ $(0.349)$ & $^{+0.402}_{-0.320}$ & $^{+0.826}_{-0.546}$ & $^{+0.960}_{-0.734}$ \\[3pt]
 $\bar{z}=0.30$ & $1.276$ $(0.367)$ & $^{+0.422}_{-0.338}$ & $^{+0.724}_{-0.556}$ & $^{+0.724}_{-0.862}$ \\[3pt]
 $\bar{z}=0.35$ & $0.741$ $(0.378)$ & $^{+0.338}_{-0.280}$ & $^{+0.754}_{-0.516}$ & $^{+1.118}_{-0.686}$ \\[3pt]
 $\bar{z}=0.39$ & $0.371$ $(0.387)$ & $^{+0.324}_{-0.262}$ & $^{+0.748}_{-0.482}$ & $^{+1.299}_{-0.666}$ \\[6pt]
 all & $0.599$ $(0.374)$ & $^{+0.134}_{-0.124}$ & $^{+0.284}_{-0.240}$ & $^{+0.450}_{-0.348}$ \\[3pt]
\hline\hline
\vspace{-30pt}
\end{tabular}
\end{table}
\begin{table}[h]
\centering
\begin{tabular}{c|cccc}
 CMASS & $\beta$ & $1\sigma$ & $2\sigma$ & $3\sigma$ \\
\hline\hline\\[-9pt]
 $\bar{r}_\void=23.7\hMpc$ & $0.459$ $(0.414)$ & $^{+0.136}_{-0.126}$ & $^{+0.284}_{-0.246}$ & $^{+0.448}_{-0.356}$ \\[3pt]
 $\bar{r}_\void=32.6\hMpc$ & $0.419$ $(0.416)$ & $^{+0.164}_{-0.152}$ & $^{+0.348}_{-0.290}$ & $^{+0.556}_{-0.418}$ \\[3pt]
 $\bar{r}_\void=41.9\hMpc$ & $0.345$ $(0.419)$ & $^{+0.184}_{-0.166}$ & $^{+0.394}_{-0.318}$ & $^{+0.630}_{-0.458}$ \\[3pt]
 $\bar{r}_\void=57.8\hMpc$ & $0.403$ $(0.423)$ & $^{+0.218}_{-0.192}$ & $^{+0.468}_{-0.366}$ & $^{+0.760}_{-0.522}$ \\[6pt]
 $\bar{z}=0.49$ & $0.431$ $(0.408)$ & $^{+0.106}_{-0.100}$ & $^{+0.222}_{-0.194}$ & $^{+0.346}_{-0.284}$ \\[3pt]
 $\bar{z}=0.53$ & $0.489$ $(0.415)$ & $^{+0.114}_{-0.108}$ & $^{+0.240}_{-0.208}$ & $^{+0.376}_{-0.304}$ \\[3pt]
 $\bar{z}=0.57$ & $0.451$ $(0.422)$ & $^{+0.116}_{-0.106}$ & $^{+0.242}_{-0.208}$ & $^{+0.378}_{-0.304}$ \\[3pt]
 $\bar{z}=0.62$ & $0.453$ $(0.429)$ & $^{+0.130}_{-0.120}$ & $^{+0.274}_{-0.232}$ & $^{+0.434}_{-0.336}$ \\[6pt]
 all & $0.457$ $(0.416)$ & $^{+0.056}_{-0.054}$ & $^{+0.116}_{-0.108}$ & $^{+0.178}_{-0.162}$ \\[3pt]
\hline\hline
\end{tabular}
\caption{Constraints on $\beta$ from various void selections in LOWZ (top) and CMASS (bottom). Theoretical values are given in parentheses, calculated as described in the caption of figure~\ref{key}.}
\label{table}
\end{table}

\begin{figure*}[!h]
\centering
\resizebox{\hsize}{!}{
\includegraphics{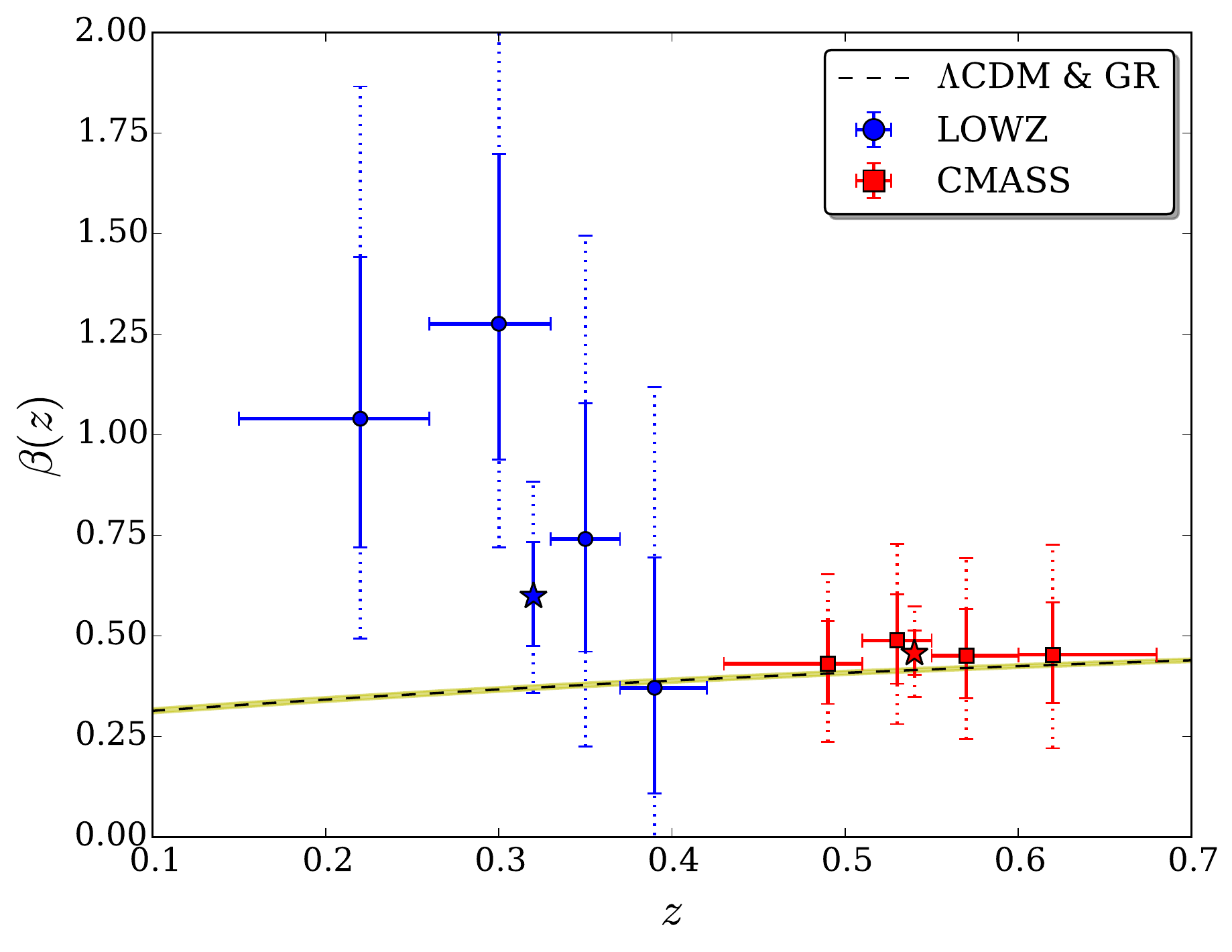}}
\caption{Growth rate constraints as a function of redshift from LOWZ (blue circles) and CMASS (red squares). Stars represent the joint constraint from voids of all redshifts in each sample. Vertical solid lines indicate $1\sigma$, dotted lines $2\sigma$ confidence intervals. Horizontal lines delineate redshift bins. The dashed line with yellow shading shows $\beta=\Omega_\mathrm{m}^{\;\gamma}(z)/b$, with $\Omega_\mathrm{m}(z=0)=0.308\pm0.012$~\cite{Planck2016}, $\gamma=0.55$~\cite{Linder2005}, and $b=1.85$~\cite{Alam2016}, assuming a flat $\Lambda$CDM cosmology and GR.}
\label{key}
\end{figure*}

\begin{figure*}[!p]
\centering
\resizebox{\hsize}{!}{
\includegraphics[width=\hsize,clip]{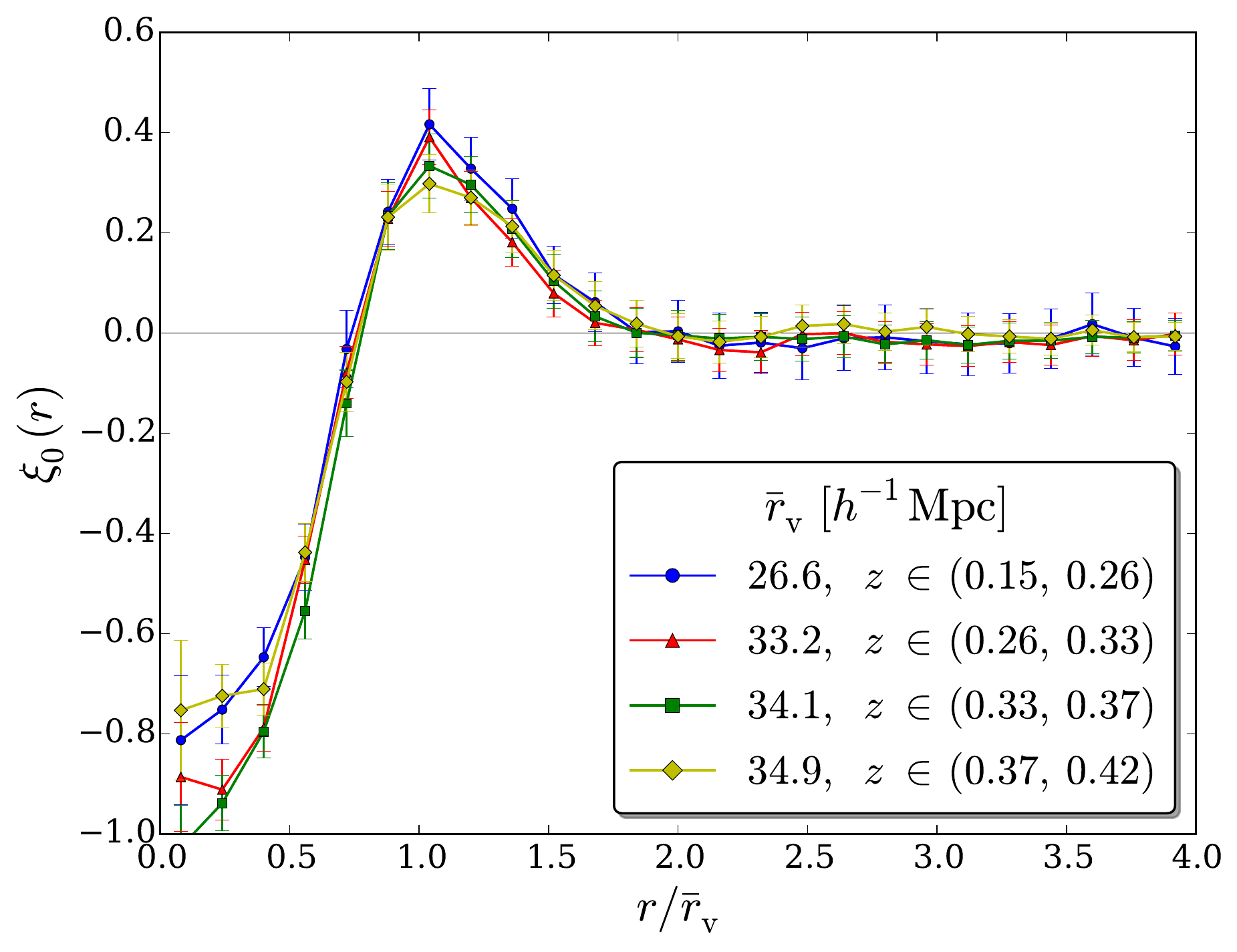}
\includegraphics[width=\hsize,clip]{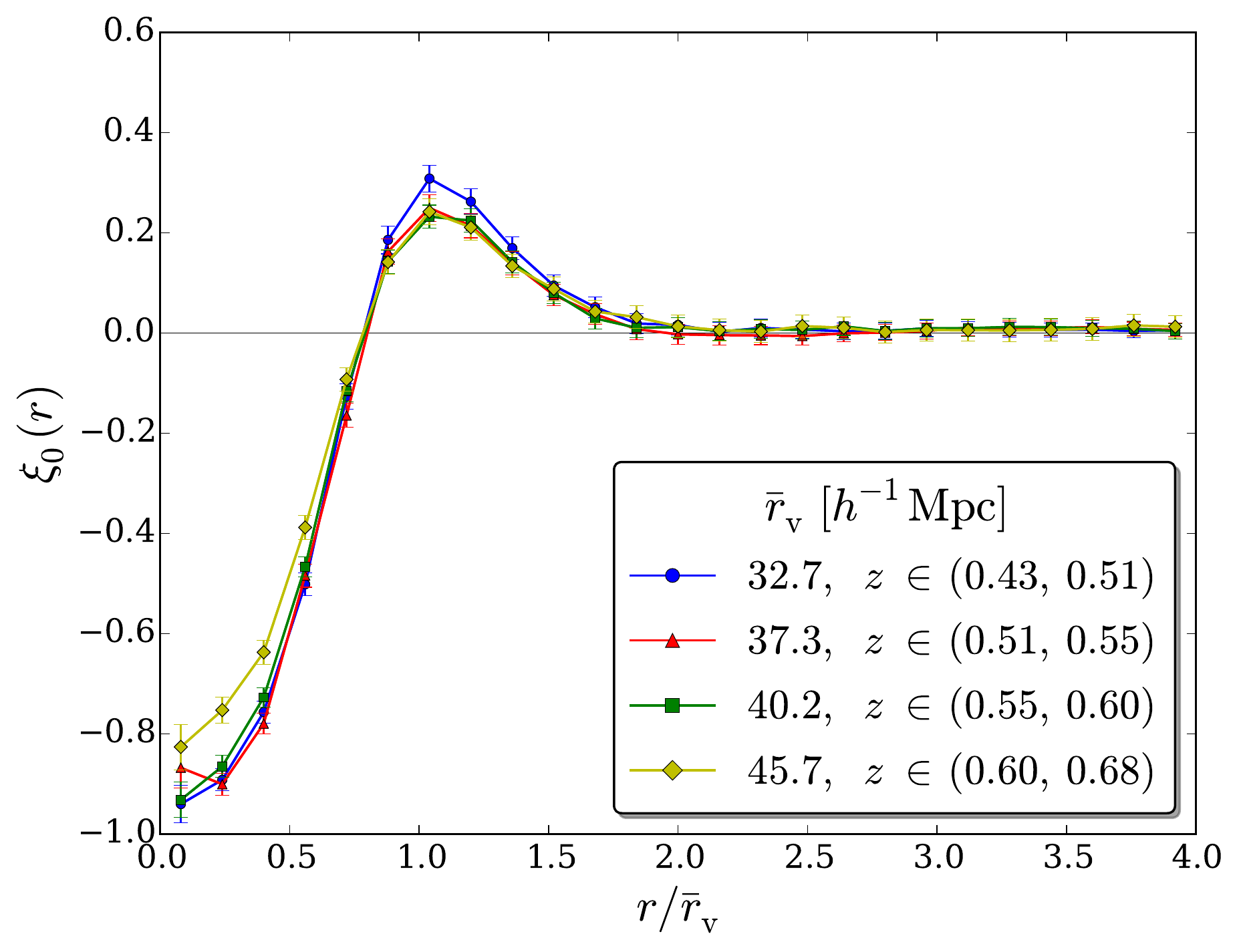}}
\resizebox{\hsize}{!}{
\includegraphics[width=\hsize,clip]{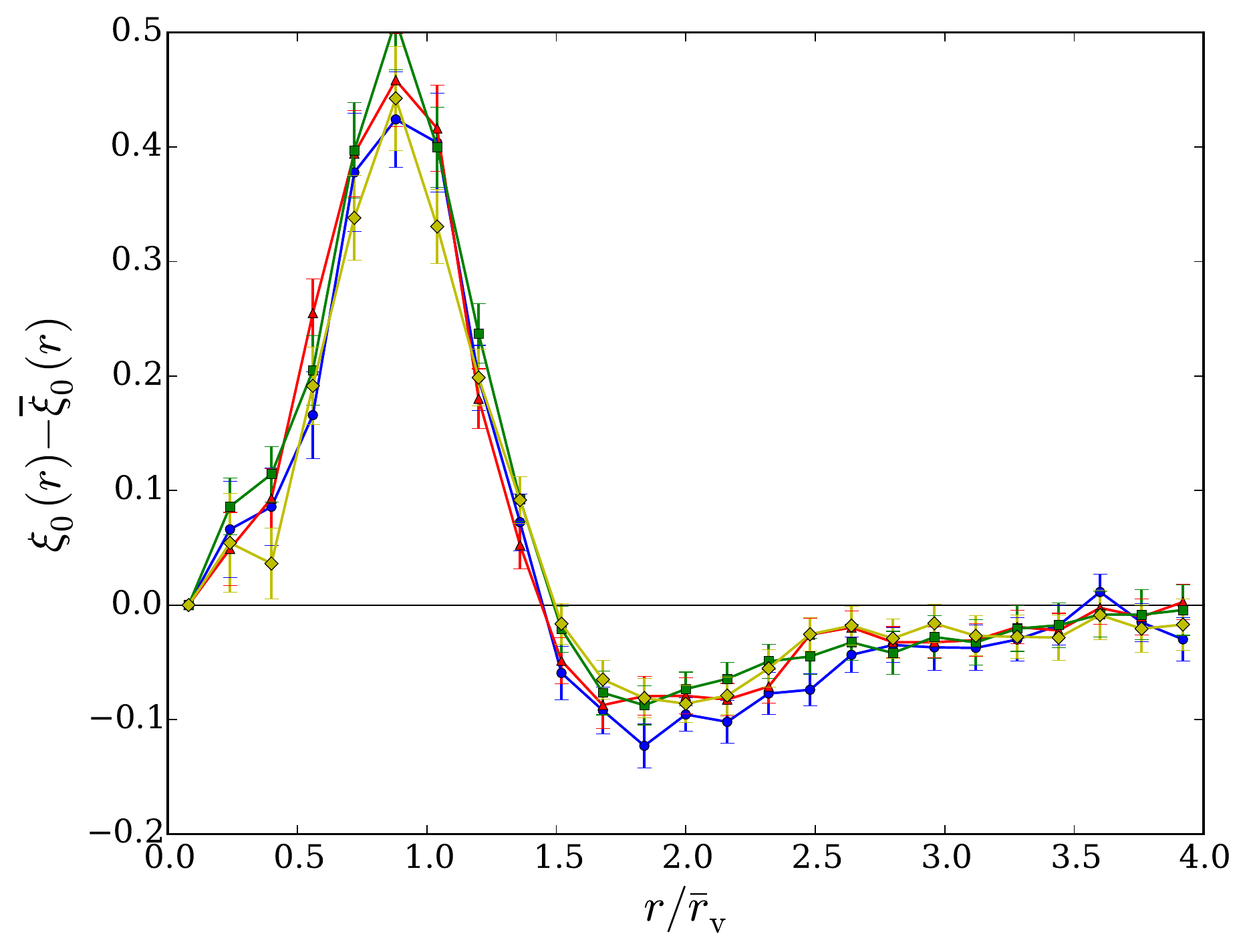}
\includegraphics[width=\hsize,clip]{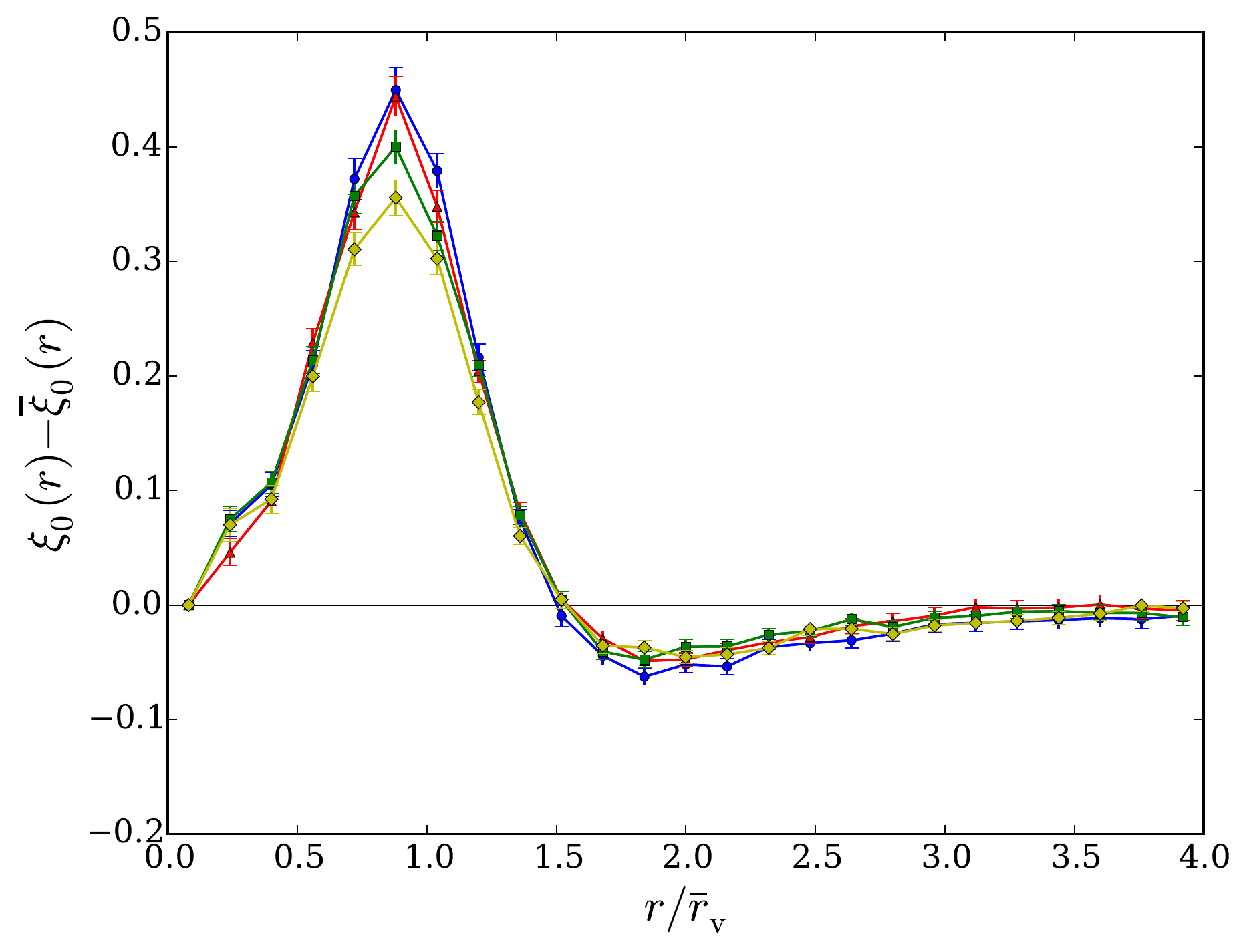}}
\resizebox{\hsize}{!}{
\includegraphics[width=\hsize,clip]{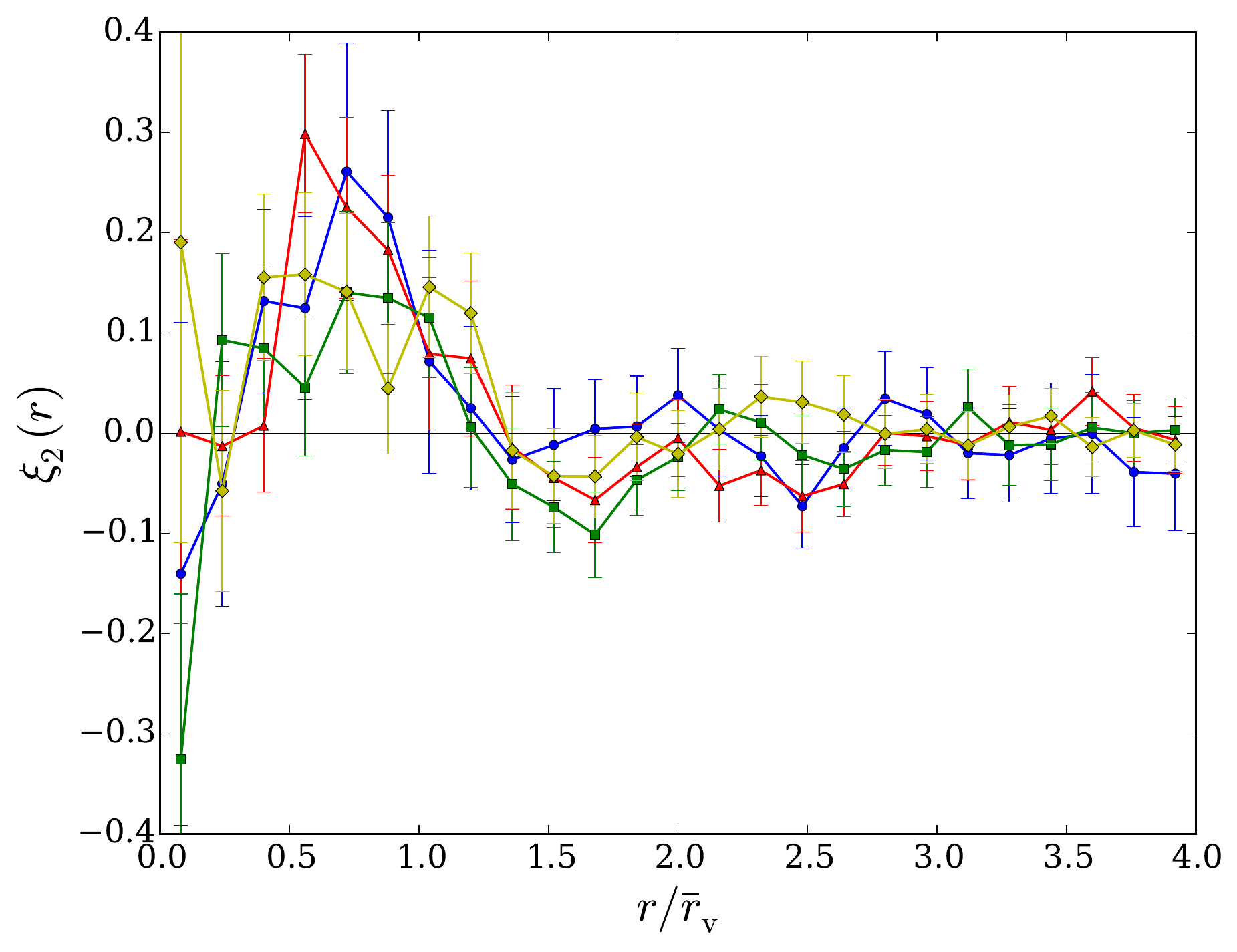}
\includegraphics[width=\hsize,clip]{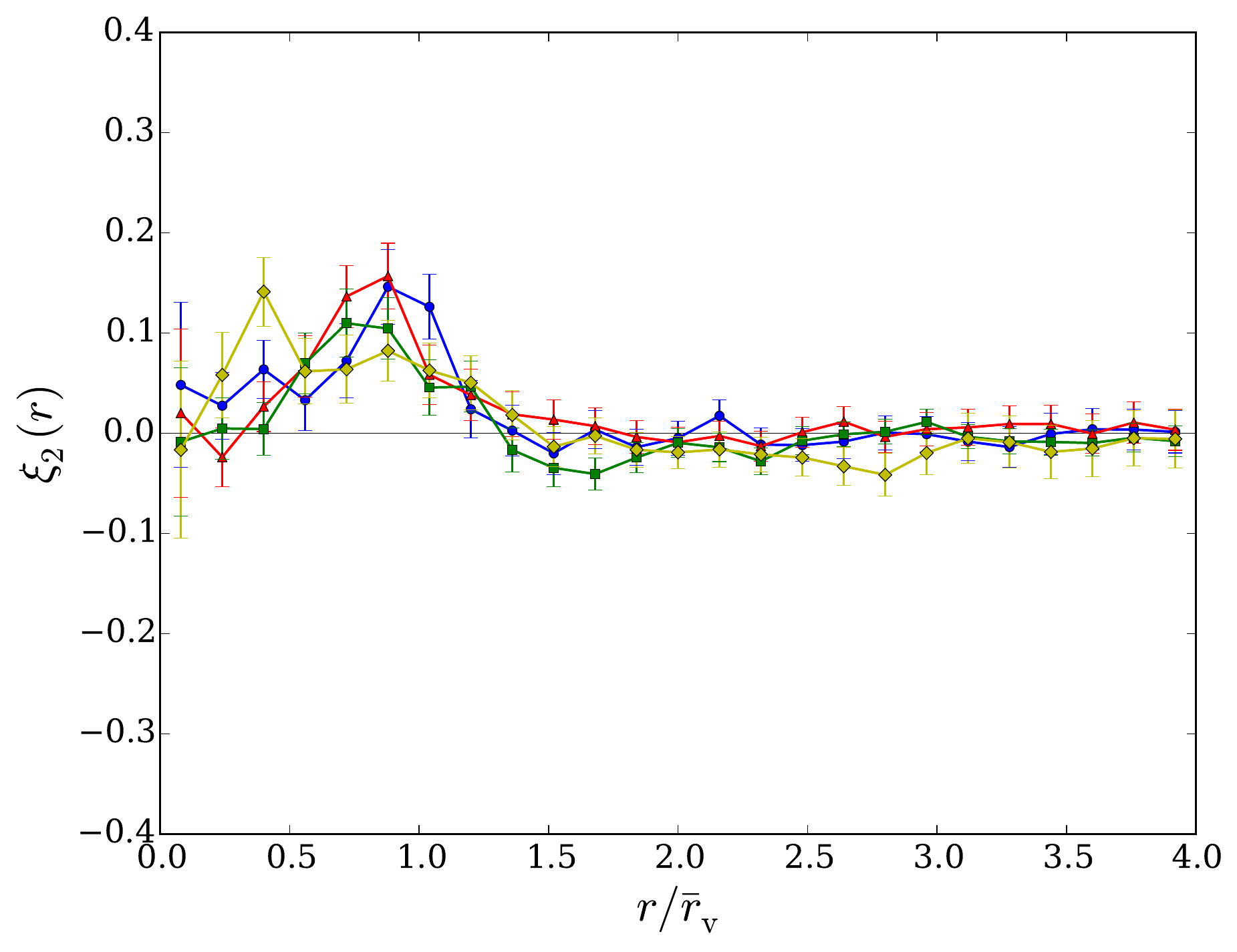}}
\caption{Multipoles of the void-galaxy cross-correlation function in LOWZ (left) and CMASS (right) for four bins in redshift. The bin edges are given in parenthesis in the figure legend.}
\label{Xvgz}
\end{figure*}

In figure~\ref{Xvgz} a similar binning has been performed in redshifts, with again roughly equal numbers of voids per bin. In this case hardly any evolution with redshift is discernible in the CMASS sample, the final constraints on $\beta$ all remain consistent to within $1\sigma$, as shown in the right panel of figure~\ref{betaz}. In contrast, the LOWZ sample does exhibit a considerable redshift evolution, suggesting higher quadrupole amplitudes and $\beta$ values towards lower redshifts. In the lowest two redshift bins, the tension between the data constraints and our model expectations clearly exceed the $2\sigma$-level, reaching about $3.2\sigma$ in the second bin. We do not observe such a behavior in the QPM mocks, as evident in figure~\ref{betaz_mock}. While this tension is still within the realm of rare statistical fluctuations, it may certainly be of interest to follow up on the observed low-redshift trend with independent data sets. One possible cause could be the redshift evolution in the linear bias $b$, which we assume as constant. However, this evolution can only account for a relative change of $\sim20\%$ in the bias at best~\cite{Marinoni2005,Clerkin2015}, and should also be present in the CMASS sample at higher redshifts. Figure~\ref{key} summarizes our $\beta$ constraints from both samples and all considered redshift bins.

Finally, one might speculate about deeper inconsistencies with our model assumptions. If indeed a higher than expected growth rate persists in the data, it could hint at an enhanced strength of gravity. Such scenarios have already been investigated with the help of simulations that incorporate force modifications due to different classes of gravity models beyond GR~\cite{Zivick2015,Cai2015,Barreira2015,Falck2017}. In such models, gravity is typically stronger in environments of low density, where screening mechanisms, such as of \emph{Chameleon} or \emph{Vainshtein} type~\cite{Clifton2012}, do not operate. This could give an explanation as to why such an effect would not have shown up already in earlier RSD analyses that only focus on the two-point statistics of galaxies, which typically trace the highly over-dense part of the cosmic web.

\section{Conclusion\label{sec:conclusion}}
We presented the first RSD multipole analysis in the observed void-galaxy cross-correlation function. It is based on public galaxy catalogs released by the BOSS collaboration, and a simple RSD model derived from linear theory. We obtain constraints on the relative growth rate $\beta$ that are competitive with state-of-the-art constraints from standard RSD analyses in the literature. For example, reference~\cite{Chuang2016b} present $\beta$ measurements with a relative accuracy of $\sim22\%$ from LOWZ and $\sim16\%$ from CMASS. In our case, these numbers are $\sim21\%$ and $\sim12\%$, respectively. We presume that common limitations in RSD analyses of the two-point clustering statistics of galaxies are mitigated in void-galaxy cross-correlations, which could explain a gain of accuracy in spite of the lower number statistics available from voids. Among these limitations, some of the most important ones are:
\begin{itemize}
 \item Nonlinear RSDs from virial motions of close-by galaxy pairs, causing the \emph{Fingers-of-God} effect~\cite{Hamaus2014c,Pisani2015,Hamaus2015,Hamaus2016}.
 \item Nonlinear clustering on small scales and at high densities~\cite{Hamaus2014a,Chan2014,Liang2016}.
 \item Nonlinear and scale-dependent galaxy bias~\cite{Pollina2017}.
 \item Impact of baryonic physics~\cite{Paillas2016}.
\end{itemize}
None of these seem to play a major role in the void-galaxy cross-correlation function in BOSS, which enables us to exploit its entire range of scales down to arbitrarily small separations between void centers and galaxies. In the binning we used, scales of slightly less than $2\hMpc$ are being probed. In contrast, the galaxy auto-correlation function is typically cut at scales below $20\hMpc$ to $40\hMpc$~\cite{Chuang2016b,SanchezA2017}, where even the most sophisticated modeling breaks down. In the data we considered, void-galaxy RSDs are perfectly described by linear theory, making the data analysis a trivial task. Above all, the quality of the signal allows us to estimate covariance matrices from the data itself, without having to rely on mock catalogs that assume a fiducial cosmology.

Reference~\cite{Cai2016} apply the same model to voids identified with a spherical under-density algorithm in the distribution of dark-matter halos from an $N$-body simulation, quoting a $9\%$ precision on $\beta$ attainable via the multipoles from a similar volume. However, they have to restrict their analysis to regions outside the void radius, claiming that scales interior to their voids are strongly affected by the velocity dispersion of galaxies. We do not observe any inconsistency with linear RSDs on scales within the effective radius of our voids. In fact, this is the regime where our signal peaks and where the model assumptions leading to equation~(\ref{xi_0_2}) are satisfied (namely concerning the pairwise bulk motion of voids). We suspect the void definition criterion in reference~\cite{Cai2016} as a potential issue, as it results in essentially empty voids with a poor sampling of tracers inside the void radius. On the other hand it leads to more prominent void ridges, which can enhance the RSD signal, but at the same time introduce potential nonlinearities.

For what concerns the CMASS sample, we find a good agreement with other growth-rate measurements in the literature. Our LOWZ analysis hints at a somewhat higher value for the growth rate than what is expected from $\Lambda$CDM and GR. In contrast, it often falls slightly below this model expectation in galaxy-clustering studies (see, e.g., figure~$15$ in reference~\cite{Alam2016}), and recently in void-galaxy cross-correlation studies at low redshift ($\bar{z}=0.052$)~\cite{Achitouv2017} and high redshift ($\bar{z}=0.727$)~\cite{Hawken2016}. The statistical significance of this finding is certainly not sufficient to claim any tension, but we consider it interesting enough to deserve further investigation.

\clearpage

\begin{acknowledgments}
N.~H. and J.~W. acknowledge support from the DFG cluster of excellence ``Origin and Structure of the Universe'' and the Trans-Regional Collaborative Research Center TRR 33 ``The Dark Universe'' of the DFG. A.~P. acknowledges financial support of the OCEVU LABEX (Grant No. ANR-11-LABX-0060) and the A*MIDEX project (Grant No. ANR-11-IDEX-0001-02) funded by the ``Investissements d'Avenir'' French government program managed by the ANR. S.~E. acknowledges financial support of the project ANR-16-CE31-0021 of the French National Research Agency (ANR).

Funding for SDSS-III~\cite{SDSS} has been provided by the Alfred P. Sloan Foundation, the Participating Institutions, the National Science Foundation, and the U.S. Department of Energy Office of Science. SDSS-III is managed by the Astrophysical Research Consortium for the Participating Institutions of the SDSS-III Collaboration including the University of Arizona, the Brazilian Participation Group, Brookhaven National Laboratory, Carnegie Mellon University, University of Florida, the French Participation Group, the German Participation Group, Harvard University, the Instituto de Astrofisica de Canarias, the Michigan State/Notre Dame/JINA Participation Group, Johns Hopkins University, Lawrence Berkeley National Laboratory, Max Planck Institute for Astrophysics, Max Planck Institute for Extraterrestrial Physics, New Mexico State University, New York University, Ohio State University, Pennsylvania State University, University of Portsmouth, Princeton University, the Spanish Participation Group, University of Tokyo, University of Utah, Vanderbilt University, University of Virginia, University of Washington, and Yale University.
\end{acknowledgments}

\bibliography{ms.bib}

\providecommand{\href}[2]{#2}\begingroup\raggedright\begin{thebibliography}{10}

\bibitem{DES}
{Dark Energy Survey (\url{http://www.darkenergysurvey.org})}.

\bibitem{EUCLID}
{Euclid (\url{http://sci.esa.int/euclid})}.

\bibitem{LSST}
{Large Synoptic Survey Telescope (\url{http://www.lsst.org})}.

\bibitem{SDSS}
{Sloan Digital Sky Survey (\url{http://www.sdss3.org})}.

\bibitem{WFIRST}
{Wide-Field Infrared Survey Telescope (\url{https://wfirst.gsfc.nasa.gov})}.

\bibitem{Peacock2001}
J.~A. {Peacock}, S.~{Cole}, P.~{Norberg}, C.~M. {Baugh}, J.~{Bland-Hawthorn},
  T.~{Bridges} et~al., \emph{{A measurement of the cosmological mass density
  from clustering in the 2dF Galaxy Redshift Survey}},
  \href{http://dx.doi.org/10.1038/35065528}{\emph{\nat} {\bfseries 410} (Mar.,
  2001) 169--173}, [\href{https://arxiv.org/abs/astro-ph/0103143}{{\ttfamily
  astro-ph/0103143}}].

\bibitem{Guzzo2008}
L.~{Guzzo}, M.~{Pierleoni}, B.~{Meneux}, E.~{Branchini}, O.~{Le F{\`e}vre},
  C.~{Marinoni} et~al., \emph{{A test of the nature of cosmic acceleration
  using galaxy redshift distortions}},
  \href{http://dx.doi.org/10.1038/nature06555}{\emph{\nat} {\bfseries 451}
  (Jan., 2008) 541--544}, [\href{https://arxiv.org/abs/0802.1944}{{\ttfamily
  0802.1944}}].

\bibitem{Satpathy2017}
S.~{Satpathy}, S.~{Alam}, S.~{Ho}, M.~{White}, N.~A. {Bahcall}, F.~{Beutler}
  et~al., \emph{{The clustering of galaxies in the completed SDSS-III Baryon
  Oscillation Spectroscopic Survey: on the measurement of growth rate using
  galaxy correlation functions}},
  \href{http://dx.doi.org/10.1093/mnras/stx883}{\emph{\mnras} {\bfseries 469}
  (Aug., 2017) 1369--1382}, [\href{https://arxiv.org/abs/1607.03148}{{\ttfamily
  1607.03148}}].

\bibitem{SanchezA2017}
A.~G. {S{\'a}nchez}, R.~{Scoccimarro}, M.~{Crocce}, J.~N. {Grieb},
  S.~{Salazar-Albornoz}, C.~{Dalla Vecchia} et~al., \emph{{The clustering of
  galaxies in the completed SDSS-III Baryon Oscillation Spectroscopic Survey:
  Cosmological implications of the configuration-space clustering wedges}},
  \href{http://dx.doi.org/10.1093/mnras/stw2443}{\emph{\mnras} {\bfseries 464}
  (Jan., 2017) 1640--1658}, [\href{https://arxiv.org/abs/1607.03147}{{\ttfamily
  1607.03147}}].

\bibitem{Beutler2017}
F.~{Beutler}, H.-J. {Seo}, S.~{Saito}, C.-H. {Chuang}, A.~J. {Cuesta}, D.~J.
  {Eisenstein} et~al., \emph{{The clustering of galaxies in the completed
  SDSS-III Baryon Oscillation Spectroscopic Survey: anisotropic galaxy
  clustering in Fourier space}},
  \href{http://dx.doi.org/10.1093/mnras/stw3298}{\emph{\mnras} {\bfseries 466}
  (Apr., 2017) 2242--2260}, [\href{https://arxiv.org/abs/1607.03150}{{\ttfamily
  1607.03150}}].

\bibitem{Grieb2017}
J.~N. {Grieb}, A.~G. {S{\'a}nchez}, S.~{Salazar-Albornoz}, R.~{Scoccimarro},
  M.~{Crocce}, C.~{Dalla Vecchia} et~al., \emph{{The clustering of galaxies in
  the completed SDSS-III Baryon Oscillation Spectroscopic Survey: Cosmological
  implications of the Fourier space wedges of the final sample}},
  \href{http://dx.doi.org/10.1093/mnras/stw3384}{\emph{\mnras} {\bfseries 467}
  (May, 2017) 2085--2112}, [\href{https://arxiv.org/abs/1607.03143}{{\ttfamily
  1607.03143}}].

\bibitem{Scoccimarro2004}
R.~{Scoccimarro}, \emph{{Redshift-space distortions, pairwise velocities, and
  nonlinearities}},
  \href{http://dx.doi.org/10.1103/PhysRevD.70.083007}{\emph{\prd} {\bfseries
  70} (Oct., 2004) 083007},
  [\href{https://arxiv.org/abs/astro-ph/0407214}{{\ttfamily
  astro-ph/0407214}}].

\bibitem{Matsubara2008}
T.~{Matsubara}, \emph{{Nonlinear perturbation theory with halo bias and
  redshift-space distortions via the Lagrangian picture}},
  \href{http://dx.doi.org/10.1103/PhysRevD.78.083519}{\emph{\prd} {\bfseries
  78} (Oct., 2008) 083519}, [\href{https://arxiv.org/abs/0807.1733}{{\ttfamily
  0807.1733}}].

\bibitem{Taruya2010}
A.~{Taruya}, T.~{Nishimichi} and S.~{Saito}, \emph{{Baryon acoustic
  oscillations in 2D: Modeling redshift-space power spectrum from perturbation
  theory}}, \href{http://dx.doi.org/10.1103/PhysRevD.82.063522}{\emph{\prd}
  {\bfseries 82} (Sept., 2010) 063522},
  [\href{https://arxiv.org/abs/1006.0699}{{\ttfamily 1006.0699}}].

\bibitem{Okumura2015}
T.~{Okumura}, N.~{Hand}, U.~{Seljak}, Z.~{Vlah} and V.~{Desjacques},
  \emph{{Galaxy power spectrum in redshift space: Combining perturbation theory
  with the halo model}},
  \href{http://dx.doi.org/10.1103/PhysRevD.92.103516}{\emph{\prd} {\bfseries
  92} (Nov., 2015) 103516}, [\href{https://arxiv.org/abs/1506.05814}{{\ttfamily
  1506.05814}}].

\bibitem{Percival2009}
W.~J. {Percival} and M.~{White}, \emph{{Testing cosmological structure
  formation using redshift-space distortions}},
  \href{http://dx.doi.org/10.1111/j.1365-2966.2008.14211.x}{\emph{\mnras}
  {\bfseries 393} (Feb., 2009) 297--308},
  [\href{https://arxiv.org/abs/0808.0003}{{\ttfamily 0808.0003}}].

\bibitem{Reid2011}
B.~A. {Reid} and M.~{White}, \emph{{Towards an accurate model of the
  redshift-space clustering of haloes in the quasi-linear regime}},
  \href{http://dx.doi.org/10.1111/j.1365-2966.2011.19379.x}{\emph{\mnras}
  {\bfseries 417} (Nov., 2011) 1913--1927},
  [\href{https://arxiv.org/abs/1105.4165}{{\ttfamily 1105.4165}}].

\bibitem{White2015}
M.~{White}, B.~{Reid}, C.-H. {Chuang}, J.~L. {Tinker}, C.~K. {McBride},
  F.~{Prada} et~al., \emph{{Tests of redshift-space distortions models in
  configuration space for the analysis of the BOSS final data release}},
  \href{http://dx.doi.org/10.1093/mnras/stu2460}{\emph{\mnras} {\bfseries 447}
  (Feb., 2015) 234--245}, [\href{https://arxiv.org/abs/1408.5435}{{\ttfamily
  1408.5435}}].

\bibitem{Shandarin2011}
S.~F. {Shandarin}, \emph{{The multi-stream flows and the dynamics of the cosmic
  web}}, \href{http://dx.doi.org/10.1088/1475-7516/2011/05/015}{\emph{\jcap}
  {\bfseries 5} (May, 2011) 015},
  [\href{https://arxiv.org/abs/1011.1924}{{\ttfamily 1011.1924}}].

\bibitem{Falck2015}
B.~{Falck} and M.~C. {Neyrinck}, \emph{{The persistent percolation of
  single-stream voids}},
  \href{http://dx.doi.org/10.1093/mnras/stv879}{\emph{\mnras} {\bfseries 450}
  (July, 2015) 3239--3253}, [\href{https://arxiv.org/abs/1410.4751}{{\ttfamily
  1410.4751}}].

\bibitem{Ramachandra2017}
N.~S. {Ramachandra} and S.~F. {Shandarin}, \emph{{Topology and geometry of the
  dark matter web: A multi-stream view}},
  \href{http://dx.doi.org/10.1093/mnras/stx183}{\emph{\mnras} {\bfseries 467}
  (May, 2017) 1748--1762}, [\href{https://arxiv.org/abs/1608.05469}{{\ttfamily
  1608.05469}}].

\bibitem{Sutter2012b}
P.~M. {Sutter}, G.~{Lavaux}, B.~D. {Wandelt} and D.~H. {Weinberg}, \emph{{A
  First Application of the Alcock-Paczynski Test to Stacked Cosmic Voids}},
  \href{http://dx.doi.org/10.1088/0004-637X/761/2/187}{\emph{\apj} {\bfseries
  761} (Dec., 2012) 187}, [\href{https://arxiv.org/abs/1208.1058}{{\ttfamily
  1208.1058}}].

\bibitem{Sutter2014b}
P.~M. {Sutter}, A.~{Pisani}, B.~D. {Wandelt} and D.~H. {Weinberg}, \emph{{A
  measurement of the Alcock-Paczy{\'n}ski effect using cosmic voids in the
  SDSS}}, \href{http://dx.doi.org/10.1093/mnras/stu1392}{\emph{\mnras}
  {\bfseries 443} (Oct., 2014) 2983--2990},
  [\href{https://arxiv.org/abs/1404.5618}{{\ttfamily 1404.5618}}].

\bibitem{Hamaus2016}
N.~{Hamaus}, A.~{Pisani}, P.~M. {Sutter}, G.~{Lavaux}, S.~{Escoffier}, B.~D.
  {Wandelt} et~al., \emph{{Constraints on Cosmology and Gravity from the
  Dynamics of Voids}},
  \href{http://dx.doi.org/10.1103/PhysRevLett.117.091302}{\emph{Physical Review
  Letters} {\bfseries 117} (Aug., 2016) 091302},
  [\href{https://arxiv.org/abs/1602.01784}{{\ttfamily 1602.01784}}].

\bibitem{Mao2017}
Q.~{Mao}, A.~A. {Berlind}, R.~J. {Scherrer}, M.~C. {Neyrinck},
  R.~{Scoccimarro}, J.~L. {Tinker} et~al., \emph{{Cosmic Voids in the SDSS DR12
  BOSS Galaxy Sample: The Alcock-Paczynski Test}},
  \href{http://dx.doi.org/10.3847/1538-4357/835/2/160}{\emph{\apj} {\bfseries
  835} (Feb., 2017) 160}, [\href{https://arxiv.org/abs/1602.06306}{{\ttfamily
  1602.06306}}].

\bibitem{Melchior2014}
P.~{Melchior}, P.~M. {Sutter}, E.~S. {Sheldon}, E.~{Krause} and B.~D.
  {Wandelt}, \emph{{First measurement of gravitational lensing by cosmic voids
  in SDSS}}, \href{http://dx.doi.org/10.1093/mnras/stu456}{\emph{\mnras}
  {\bfseries 440} (June, 2014) 2922--2927},
  [\href{https://arxiv.org/abs/1309.2045}{{\ttfamily 1309.2045}}].

\bibitem{Clampitt2015}
J.~{Clampitt} and B.~{Jain}, \emph{{Lensing measurements of the mass
  distribution in SDSS voids}},
  \href{http://dx.doi.org/10.1093/mnras/stv2215}{\emph{\mnras} {\bfseries 454}
  (Dec., 2015) 3357--3365}, [\href{https://arxiv.org/abs/1404.1834}{{\ttfamily
  1404.1834}}].

\bibitem{SanchezC2017}
C.~{S{\'a}nchez}, J.~{Clampitt}, A.~{Kovacs}, B.~{Jain},
  J.~{Garc{\'{\i}}a-Bellido}, S.~{Nadathur} et~al., \emph{{Cosmic voids and
  void lensing in the Dark Energy Survey Science Verification data}},
  \href{http://dx.doi.org/10.1093/mnras/stw2745}{\emph{\mnras} {\bfseries 465}
  (Feb., 2017) 746--759}, [\href{https://arxiv.org/abs/1605.03982}{{\ttfamily
  1605.03982}}].

\bibitem{Chantavat2017}
T.~{Chantavat}, U.~{Sawangwit} and B.~D. {Wandelt}, \emph{{Void Profile from
  Planck Lensing Potential Map}},
  \href{http://dx.doi.org/10.3847/1538-4357/836/2/156}{\emph{\apj} {\bfseries
  836} (Feb., 2017) 156}, [\href{https://arxiv.org/abs/1702.01009}{{\ttfamily
  1702.01009}}].

\bibitem{Granett2008}
B.~R. {Granett}, M.~C. {Neyrinck} and I.~{Szapudi}, \emph{{An Imprint of
  Superstructures on the Microwave Background due to the Integrated Sachs-Wolfe
  Effect}}, \href{http://dx.doi.org/10.1086/591670}{\emph{\apjl} {\bfseries
  683} (Aug., 2008) L99}, [\href{https://arxiv.org/abs/0805.3695}{{\ttfamily
  0805.3695}}].

\bibitem{Nadathur2016}
S.~{Nadathur} and R.~{Crittenden}, \emph{{A Detection of the Integrated
  Sachs-Wolfe Imprint of Cosmic Superstructures Using a Matched-filter
  Approach}}, \href{http://dx.doi.org/10.3847/2041-8205/830/1/L19}{\emph{\apjl}
  {\bfseries 830} (Oct., 2016) L19},
  [\href{https://arxiv.org/abs/1608.08638}{{\ttfamily 1608.08638}}].

\bibitem{Cai2017}
Y.-C. {Cai}, M.~{Neyrinck}, Q.~{Mao}, J.~A. {Peacock}, I.~{Szapudi} and A.~A.
  {Berlind}, \emph{{The lensing and temperature imprints of voids on the cosmic
  microwave background}},
  \href{http://dx.doi.org/10.1093/mnras/stw3299}{\emph{\mnras} {\bfseries 466}
  (Apr., 2017) 3364--3375}, [\href{https://arxiv.org/abs/1609.00301}{{\ttfamily
  1609.00301}}].

\bibitem{Kovacs2017}
A.~{Kov{\'a}cs}, C.~{S{\'a}nchez}, J.~{Garc{\'{\i}}a-Bellido}, S.~{Nadathur},
  R.~{Crittenden}, D.~{Gruen} et~al., \emph{{Imprint of DES superstructures on
  the cosmic microwave background}},
  \href{http://dx.doi.org/10.1093/mnras/stw2968}{\emph{\mnras} {\bfseries 465}
  (Mar., 2017) 4166--4179}, [\href{https://arxiv.org/abs/1610.00637}{{\ttfamily
  1610.00637}}].

\bibitem{Kitaura2016}
F.-S. {Kitaura}, C.-H. {Chuang}, Y.~{Liang}, C.~{Zhao}, C.~{Tao},
  S.~{Rodr{\'{\i}}guez-Torres} et~al., \emph{{Signatures of the Primordial
  Universe from Its Emptiness: Measurement of Baryon Acoustic Oscillations from
  Minima of the Density Field}},
  \href{http://dx.doi.org/10.1103/PhysRevLett.116.171301}{\emph{Physical Review
  Letters} {\bfseries 116} (Apr., 2016) 171301},
  [\href{https://arxiv.org/abs/1511.04405}{{\ttfamily 1511.04405}}].

\bibitem{Paz2013}
D.~{Paz}, M.~{Lares}, L.~{Ceccarelli}, N.~{Padilla} and D.~G. {Lambas},
  \emph{{Clues on void evolution-II. Measuring density and velocity profiles on
  SDSS galaxy redshift space distortions}},
  \href{http://dx.doi.org/10.1093/mnras/stt1836}{\emph{\mnras} {\bfseries 436}
  (Dec., 2013) 3480--3491}, [\href{https://arxiv.org/abs/1306.5799}{{\ttfamily
  1306.5799}}].

\bibitem{Achitouv2017}
I.~{Achitouv}, C.~{Blake}, P.~{Carter}, J.~{Koda} and F.~{Beutler},
  \emph{{Consistency of the growth rate in different environments with the
  6-degree Field Galaxy Survey: Measurement of the void-galaxy and
  galaxy-galaxy correlation functions}},
  \href{http://dx.doi.org/10.1103/PhysRevD.95.083502}{\emph{\prd} {\bfseries
  95} (Apr., 2017) 083502}, [\href{https://arxiv.org/abs/1606.03092}{{\ttfamily
  1606.03092}}].

\bibitem{Hawken2016}
A.~J. {Hawken}, B.~R. {Granett}, A.~{Iovino}, L.~{Guzzo}, J.~A. {Peacock},
  S.~{de la Torre} et~al., \emph{{The VIMOS Public Extragalactic Redshift
  Survey: Measuring the growth rate of structure around cosmic voids}},
  {\emph{ArXiv e-prints} (Nov., 2016) },
  [\href{https://arxiv.org/abs/1611.07046}{{\ttfamily 1611.07046}}].

\bibitem{Kaiser1987}
N.~{Kaiser}, \emph{{Clustering in real space and in redshift space}},
  \href{http://dx.doi.org/10.1093/mnras/227.1.1}{\emph{\mnras} {\bfseries 227}
  (July, 1987) 1--21}.

\bibitem{Hamilton1998}
A.~J.~S. {Hamilton}, \emph{{Linear Redshift Distortions: a Review}},  in
  \emph{The Evolving Universe} (D.~{Hamilton}, ed.), vol.~231 of
  \emph{Astrophysics and Space Science Library}, p.~185, 1998.
\newblock \href{https://arxiv.org/abs/astro-ph/9708102}{{\ttfamily
  astro-ph/9708102}}.
\newblock \href{http://dx.doi.org/10.1007/978-94-011-4960-0_17}{DOI}.

\bibitem{Sutter2014c}
P.~M. {Sutter}, P.~{Elahi}, B.~{Falck}, J.~{Onions}, N.~{Hamaus}, A.~{Knebe}
  et~al., \emph{{The life and death of cosmic voids}},
  \href{http://dx.doi.org/10.1093/mnras/stu1845}{\emph{\mnras} {\bfseries 445}
  (Dec., 2014) 1235--1244}, [\href{https://arxiv.org/abs/1403.7525}{{\ttfamily
  1403.7525}}].

\bibitem{Lambas2016}
D.~G. {Lambas}, M.~{Lares}, L.~{Ceccarelli}, A.~N. {Ruiz}, D.~J. {Paz}, V.~E.
  {Maldonado} et~al., \emph{{The sparkling Universe: the coherent motions of
  cosmic voids}}, \href{http://dx.doi.org/10.1093/mnrasl/slv151}{\emph{\mnras}
  {\bfseries 455} (Jan., 2016) L99--L103},
  [\href{https://arxiv.org/abs/1510.00712}{{\ttfamily 1510.00712}}].

\bibitem{Chuang2016a}
C.-H. {Chuang}, F.-S. {Kitaura}, Y.~{Liang}, A.~{Font-Ribera}, C.~{Zhao},
  P.~{McDonald} et~al., \emph{{Linear redshift space distortions for cosmic
  voids based on galaxies in redshift space}},
  \href{http://dx.doi.org/10.1103/PhysRevD.95.063528}{\emph{\prd} {\bfseries
  95} (Mar., 2017) 063528}, [\href{https://arxiv.org/abs/1605.05352}{{\ttfamily
  1605.05352}}].

\bibitem{Peebles1980}
P.~J.~E. {Peebles}, \emph{{The large-scale structure of the universe}}.
\newblock Princeton University Press, Princeton U.S.A., (1980).

\bibitem{Hamaus2014b}
N.~{Hamaus}, P.~M. {Sutter} and B.~D. {Wandelt}, \emph{{Universal Density
  Profile for Cosmic Voids}},
  \href{http://dx.doi.org/10.1103/PhysRevLett.112.251302}{\emph{Physical Review
  Letters} {\bfseries 112} (June, 2014) 251302},
  [\href{https://arxiv.org/abs/1403.5499}{{\ttfamily 1403.5499}}].

\bibitem{Linder2005}
E.~V. {Linder}, \emph{{Cosmic growth history and expansion history}},
  \href{http://dx.doi.org/10.1103/PhysRevD.72.043529}{\emph{\prd} {\bfseries
  72} (Aug., 2005) 043529},
  [\href{https://arxiv.org/abs/astro-ph/0507263}{{\ttfamily
  astro-ph/0507263}}].

\bibitem{Leclercq2015}
F.~{Leclercq}, J.~{Jasche}, P.~M. {Sutter}, N.~{Hamaus} and B.~{Wandelt},
  \emph{{Dark matter voids in the SDSS galaxy survey}},
  \href{http://dx.doi.org/10.1088/1475-7516/2015/03/047}{\emph{\jcap}
  {\bfseries 3} (Mar., 2015) 047},
  [\href{https://arxiv.org/abs/1410.0355}{{\ttfamily 1410.0355}}].

\bibitem{Pollina2017}
G.~{Pollina}, N.~{Hamaus}, K.~{Dolag}, J.~{Weller}, M.~{Baldi} and
  L.~{Moscardini}, \emph{{On the linearity of tracer bias around voids}},
  \href{http://dx.doi.org/10.1093/mnras/stx785}{\emph{\mnras} {\bfseries 469}
  (July, 2017) 787--799}, [\href{https://arxiv.org/abs/1610.06176}{{\ttfamily
  1610.06176}}].

\bibitem{Cai2016}
Y.-C. {Cai}, A.~{Taylor}, J.~A. {Peacock} and N.~{Padilla},
  \emph{{Redshift-space distortions around voids}},
  \href{http://dx.doi.org/10.1093/mnras/stw1809}{\emph{\mnras} {\bfseries 462}
  (Nov., 2016) 2465--2477}, [\href{https://arxiv.org/abs/1603.05184}{{\ttfamily
  1603.05184}}].

\bibitem{Eisenstein2011}
D.~J. {Eisenstein}, D.~H. {Weinberg}, E.~{Agol}, H.~{Aihara}, C.~{Allende
  Prieto}, S.~F. {Anderson} et~al., \emph{{SDSS-III: Massive Spectroscopic
  Surveys of the Distant Universe, the Milky Way, and Extra-Solar Planetary
  Systems}}, \href{http://dx.doi.org/10.1088/0004-6256/142/3/72}{\emph{\aj}
  {\bfseries 142} (Sept., 2011) 72},
  [\href{https://arxiv.org/abs/1101.1529}{{\ttfamily 1101.1529}}].

\bibitem{Dawson2013}
K.~S. {Dawson}, D.~J. {Schlegel}, C.~P. {Ahn}, S.~F. {Anderson},
  {\'E}.~{Aubourg}, S.~{Bailey} et~al., \emph{{The Baryon Oscillation
  Spectroscopic Survey of SDSS-III}},
  \href{http://dx.doi.org/10.1088/0004-6256/145/1/10}{\emph{\aj} {\bfseries
  145} (Jan., 2013) 10}, [\href{https://arxiv.org/abs/1208.0022}{{\ttfamily
  1208.0022}}].

\bibitem{Reid2016}
B.~{Reid}, S.~{Ho}, N.~{Padmanabhan}, W.~J. {Percival}, J.~{Tinker},
  R.~{Tojeiro} et~al., \emph{{SDSS-III Baryon Oscillation Spectroscopic Survey
  Data Release 12: galaxy target selection and large-scale structure
  catalogues}}, \href{http://dx.doi.org/10.1093/mnras/stv2382}{\emph{\mnras}
  {\bfseries 455} (Jan., 2016) 1553--1573},
  [\href{https://arxiv.org/abs/1509.06529}{{\ttfamily 1509.06529}}].

\bibitem{Alam2016}
S.~{Alam}, M.~{Ata}, S.~{Bailey}, F.~{Beutler}, D.~{Bizyaev}, J.~A. {Blazek}
  et~al., \emph{{The clustering of galaxies in the completed SDSS-III Baryon
  Oscillation Spectroscopic Survey: cosmological analysis of the DR12 galaxy
  sample}}, {\emph{ArXiv e-prints} (July, 2016) },
  [\href{https://arxiv.org/abs/1607.03155}{{\ttfamily 1607.03155}}].

\bibitem{White2014}
M.~{White}, J.~L. {Tinker} and C.~K. {McBride}, \emph{{Mock galaxy catalogues
  using the quick particle mesh method}},
  \href{http://dx.doi.org/10.1093/mnras/stt2071}{\emph{\mnras} {\bfseries 437}
  (Jan., 2014) 2594--2606}, [\href{https://arxiv.org/abs/1309.5532}{{\ttfamily
  1309.5532}}].

\bibitem{GilMarin2017}
H.~{Gil-Mar{\'{\i}}n}, W.~J. {Percival}, L.~{Verde}, J.~R. {Brownstein}, C.-H.
  {Chuang}, F.-S. {Kitaura} et~al., \emph{{The clustering of galaxies in the
  SDSS-III Baryon Oscillation Spectroscopic Survey: RSD measurement from the
  power spectrum and bispectrum of the DR12 BOSS galaxies}},
  \href{http://dx.doi.org/10.1093/mnras/stw2679}{\emph{\mnras} {\bfseries 465}
  (Feb., 2017) 1757--1788}, [\href{https://arxiv.org/abs/1606.00439}{{\ttfamily
  1606.00439}}].

\bibitem{Sutter2015}
P.~M. {Sutter}, G.~{Lavaux}, N.~{Hamaus}, A.~{Pisani}, B.~D. {Wandelt},
  M.~{Warren} et~al., \emph{{VIDE: The Void IDentification and Examination
  toolkit}},
  \href{http://dx.doi.org/10.1016/j.ascom.2014.10.002}{\emph{Astronomy and
  Computing} {\bfseries 9} (Mar., 2015) 1--9},
  [\href{https://arxiv.org/abs/1406.1191}{{\ttfamily 1406.1191}}].

\bibitem{Neyrinck2008}
M.~C. {Neyrinck}, \emph{{ZOBOV: a parameter-free void-finding algorithm}},
  \href{http://dx.doi.org/10.1111/j.1365-2966.2008.13180.x}{\emph{\mnras}
  {\bfseries 386} (June, 2008) 2101--2109},
  [\href{https://arxiv.org/abs/0712.3049}{{\ttfamily 0712.3049}}].

\bibitem{Platen2007}
E.~{Platen}, R.~{van de Weygaert} and B.~J.~T. {Jones}, \emph{{A cosmic
  watershed: the WVF void detection technique}},
  \href{http://dx.doi.org/10.1111/j.1365-2966.2007.12125.x}{\emph{\mnras}
  {\bfseries 380} (Sept., 2007) 551--570},
  [\href{https://arxiv.org/abs/0706.2788}{{\ttfamily 0706.2788}}].

\bibitem{Sutter2012a}
P.~M. {Sutter}, G.~{Lavaux}, B.~D. {Wandelt} and D.~H. {Weinberg}, \emph{{A
  Public Void Catalog from the SDSS DR7 Galaxy Redshift Surveys Based on the
  Watershed Transform}},
  \href{http://dx.doi.org/10.1088/0004-637X/761/1/44}{\emph{\apj} {\bfseries
  761} (Dec., 2012) 44}, [\href{https://arxiv.org/abs/1207.2524}{{\ttfamily
  1207.2524}}].

\bibitem{Hamaus2015}
N.~{Hamaus}, P.~M. {Sutter}, G.~{Lavaux} and B.~D. {Wandelt}, \emph{{Probing
  cosmology and gravity with redshift-space distortions around voids}},
  \href{http://dx.doi.org/10.1088/1475-7516/2015/11/036}{\emph{\jcap}
  {\bfseries 11} (Nov., 2015) 036},
  [\href{https://arxiv.org/abs/1507.04363}{{\ttfamily 1507.04363}}].

\bibitem{Landy1993}
S.~D. {Landy} and A.~S. {Szalay}, \emph{{Bias and variance of angular
  correlation functions}}, \href{http://dx.doi.org/10.1086/172900}{\emph{\apj}
  {\bfseries 412} (July, 1993) 64--71}.

\bibitem{Feldman1994}
H.~A. {Feldman}, N.~{Kaiser} and J.~A. {Peacock}, \emph{{Power-spectrum
  analysis of three-dimensional redshift surveys}},
  \href{http://dx.doi.org/10.1086/174036}{\emph{\apj} {\bfseries 426} (May,
  1994) 23--37}, [\href{https://arxiv.org/abs/astro-ph/9304022}{{\ttfamily
  astro-ph/9304022}}].

\bibitem{Sutter2014a}
P.~M. {Sutter}, G.~{Lavaux}, N.~{Hamaus}, B.~D. {Wandelt}, D.~H. {Weinberg} and
  M.~S. {Warren}, \emph{{Sparse sampling, galaxy bias, and voids}},
  \href{http://dx.doi.org/10.1093/mnras/stu893}{\emph{\mnras} {\bfseries 442}
  (July, 2014) 462--471}, [\href{https://arxiv.org/abs/1309.5087}{{\ttfamily
  1309.5087}}].

\bibitem{Marinoni2005}
C.~{Marinoni}, O.~{Le F{\`e}vre}, B.~{Meneux}, A.~{Iovino}, A.~{Pollo},
  O.~{Ilbert} et~al., \emph{{The VIMOS VLT Deep Survey. Evolution of the
  non-linear galaxy bias up to z = 1.5}},
  \href{http://dx.doi.org/10.1051/0004-6361:20052966}{\emph{\aap} {\bfseries
  442} (Nov., 2005) 801--825},
  [\href{https://arxiv.org/abs/astro-ph/0506561}{{\ttfamily
  astro-ph/0506561}}].

\bibitem{Clerkin2015}
L.~{Clerkin}, D.~{Kirk}, O.~{Lahav}, F.~B. {Abdalla} and E.~{Gazta{\~n}aga},
  \emph{{A prescription for galaxy biasing evolution as a nuisance parameter}},
  \href{http://dx.doi.org/10.1093/mnras/stu2754}{\emph{\mnras} {\bfseries 448}
  (Apr., 2015) 1389--1401}, [\href{https://arxiv.org/abs/1405.5521}{{\ttfamily
  1405.5521}}].

\bibitem{Zivick2015}
P.~{Zivick}, P.~M. {Sutter}, B.~D. {Wandelt}, B.~{Li} and T.~Y. {Lam},
  \emph{{Using cosmic voids to distinguish f(R) gravity in future galaxy
  surveys}}, \href{http://dx.doi.org/10.1093/mnras/stv1209}{\emph{\mnras}
  {\bfseries 451} (Aug., 2015) 4215--4222},
  [\href{https://arxiv.org/abs/1411.5694}{{\ttfamily 1411.5694}}].

\bibitem{Cai2015}
Y.-C. {Cai}, N.~{Padilla} and B.~{Li}, \emph{{Testing gravity using cosmic
  voids}}, \href{http://dx.doi.org/10.1093/mnras/stv777}{\emph{\mnras}
  {\bfseries 451} (July, 2015) 1036--1055},
  [\href{https://arxiv.org/abs/1410.1510}{{\ttfamily 1410.1510}}].

\bibitem{Barreira2015}
A.~{Barreira}, M.~{Cautun}, B.~{Li}, C.~M. {Baugh} and S.~{Pascoli},
  \emph{{Weak lensing by voids in modified lensing potentials}},
  \href{http://dx.doi.org/10.1088/1475-7516/2015/08/028}{\emph{\jcap}
  {\bfseries 8} (Aug., 2015) 028},
  [\href{https://arxiv.org/abs/1505.05809}{{\ttfamily 1505.05809}}].

\bibitem{Falck2017}
B.~{Falck}, K.~{Koyama}, G.-b. {Zhao} and M.~{Cautun}, \emph{{Using Voids to
  Unscreen Modified Gravity}}, {\emph{ArXiv e-prints} (Apr., 2017) },
  [\href{https://arxiv.org/abs/1704.08942}{{\ttfamily 1704.08942}}].

\bibitem{Clifton2012}
T.~{Clifton}, P.~G. {Ferreira}, A.~{Padilla} and C.~{Skordis}, \emph{{Modified
  gravity and cosmology}},
  \href{http://dx.doi.org/10.1016/j.physrep.2012.01.001}{\emph{\physrep}
  {\bfseries 513} (Mar., 2012) 1--189},
  [\href{https://arxiv.org/abs/1106.2476}{{\ttfamily 1106.2476}}].

\bibitem{Planck2016}
{Planck Collaboration}, P.~A.~R. {Ade}, N.~{Aghanim}, M.~{Arnaud},
  M.~{Ashdown}, J.~{Aumont} et~al., \emph{{Planck 2015 results. XIII.
  Cosmological parameters}},
  \href{http://dx.doi.org/10.1051/0004-6361/201525830}{\emph{\aap} {\bfseries
  594} (Sept., 2016) A13}, [\href{https://arxiv.org/abs/1502.01589}{{\ttfamily
  1502.01589}}].

\bibitem{Chuang2016b}
C.-H. {Chuang}, M.~{Pellejero-Ibanez}, S.~{Rodr{\'{\i}}guez-Torres}, A.~J.
  {Ross}, G.-b. {Zhao}, Y.~{Wang} et~al., \emph{{The Clustering of Galaxies in
  the Completed SDSS-III Baryon Oscillation Spectroscopic Survey: single-probe
  measurements from DR12 galaxy clustering -- towards an accurate model}},
  {\emph{ArXiv e-prints} (July, 2016) },
  [\href{https://arxiv.org/abs/1607.03151}{{\ttfamily 1607.03151}}].

\bibitem{Hamaus2014c}
N.~{Hamaus}, P.~M. {Sutter}, G.~{Lavaux} and B.~D. {Wandelt}, \emph{{Testing
  cosmic geometry without dynamic distortions using voids}},
  \href{http://dx.doi.org/10.1088/1475-7516/2014/12/013}{\emph{\jcap}
  {\bfseries 12} (Dec., 2014) 013},
  [\href{https://arxiv.org/abs/1409.3580}{{\ttfamily 1409.3580}}].

\bibitem{Pisani2015}
A.~{Pisani}, P.~M. {Sutter} and B.~D. {Wandelt}, \emph{{Mastering the effects
  of peculiar velocities in cosmic voids}}, {\emph{ArXiv e-prints} (June, 2015)
  }, [\href{https://arxiv.org/abs/1506.07982}{{\ttfamily 1506.07982}}].

\bibitem{Hamaus2014a}
N.~{Hamaus}, B.~D. {Wandelt}, P.~M. {Sutter}, G.~{Lavaux} and M.~S. {Warren},
  \emph{{Cosmology with Void-Galaxy Correlations}},
  \href{http://dx.doi.org/10.1103/PhysRevLett.112.041304}{\emph{Physical Review
  Letters} {\bfseries 112} (Jan., 2014) 041304},
  [\href{https://arxiv.org/abs/1307.2571}{{\ttfamily 1307.2571}}].

\bibitem{Chan2014}
K.~C. {Chan}, N.~{Hamaus} and V.~{Desjacques}, \emph{{Large-scale clustering of
  cosmic voids}},
  \href{http://dx.doi.org/10.1103/PhysRevD.90.103521}{\emph{\prd} {\bfseries
  90} (Nov., 2014) 103521}, [\href{https://arxiv.org/abs/1409.3849}{{\ttfamily
  1409.3849}}].

\bibitem{Liang2016}
Y.~{Liang}, C.~{Zhao}, C.-H. {Chuang}, F.-S. {Kitaura} and C.~{Tao},
  \emph{{Measuring baryon acoustic oscillations from the clustering of voids}},
  \href{http://dx.doi.org/10.1093/mnras/stw884}{\emph{\mnras} {\bfseries 459}
  (July, 2016) 4020--4028}, [\href{https://arxiv.org/abs/1511.04391}{{\ttfamily
  1511.04391}}].

\bibitem{Paillas2016}
E.~{Paillas}, C.~D.~P. {Lagos}, N.~{Padilla}, P.~{Tissera}, J.~{Helly} and
  M.~{Schaller}, \emph{{Baryon effects on void statistics in the EAGLE
  simulation}}, {\emph{ArXiv e-prints} (Aug., 2016) },
  [\href{https://arxiv.org/abs/1609.00101}{{\ttfamily 1609.00101}}].

\end{thebibliography}\endgroup
\bibliographystyle{JHEP.bst}

\end{document}